\documentclass[a4paper,english,twoside,openright,11pt]{article}

\usepackage[latin1]{inputenc} 
\usepackage[T1]{fontenc}
\usepackage[safe]{textcomp}
\usepackage{lmodern} 
\usepackage{epsfig}
\usepackage{float}
\usepackage{amstext}  
\usepackage{amsfonts}
\usepackage{amssymb} 
\usepackage{amsmath}
\usepackage{amsthm}
\usepackage{appendix}
\usepackage{bm}   
\usepackage{listings}
\usepackage{enumerate}     
\usepackage[bottom]{footmisc} 
\usepackage{array}           
\usepackage{parskip}        
\usepackage{xcolor} 
\usepackage{color}
\usepackage{framed}
\usepackage[hyphens]{url} 
\usepackage[Sonny]{fncychap}
\usepackage{fancyhdr}
\usepackage{subfig}
\usepackage[raggedright]{sidecap}
\usepackage[plainpages=false]{hyperref}
\usepackage{cite}
\usepackage{slashed}
\usepackage{tikz}
\usepackage[absolute]{textpos}
\usepackage{pstricks}
\usepackage{axodraw4j}
\usepackage{graphicx}
\usepackage{dsfont}
\usepackage{multirow}
\usepackage{booktabs}
\makeindex

\newcommand{\p}{\partial}

\newcommand{\f}[2]{\frac{#1}{#2}}
\newcommand{\sss}[1]{\scriptscriptstyle#1}
\newcommand{\vv}[2]{\left( \begin{array}{c} #1 \\ #2  \end{array} \right)}
\newcommand{\bea}{\begin{eqnarray}}
\newcommand{\eea}{\end{eqnarray}}
\newcommand{\be}{\begin{equation}}
\newcommand{\ee}{\end{equation}}
\newcommand{\ba}{\begin{align}}
\newcommand{\ea}{\end{align}}
\newcommand{\beas}{\begin{eqnarray*}}
\newcommand{\eeas}{\end{eqnarray*}}
\newcommand{\bes}{\begin{equation*}}
\newcommand{\ees}{\end{equation*}}
\newcommand{\bas}{\begin{align*}}
\newcommand{\eas}{\end{align*}}

\newcommand{\ssL}{{\mathcal L}} 
\newcommand{\eps}{{\varepsilon}}
\newcommand{\cd}{{\cdot}} 
\newcommand{\cf}{C_{\scriptscriptstyle{F}}} 
\newcommand{\ca}{C_{\scriptscriptstyle{A}}}
\newcommand{\tr}{T_{\scriptscriptstyle{F}}}

\newcommand{\dR}{d_{\scriptscriptstyle{R}}}

\newcommand{\Nl}{n_{\scriptscriptstyle{l}}}

\newcommand{\Nf}{n_{\scriptscriptstyle{f}}}
\newcommand{\gs}{g_{\scriptscriptstyle{s}}}
\newcommand{\yt}{y_{\scriptscriptstyle{t}}}

\newcommand{\als}{\alpha_{\scriptscriptstyle{s}}}

\newcommand{\lb}{\left(}
\newcommand{\rb}{\right)}

\definecolor{bluemar}{rgb}{0,0,.5}
\definecolor{redmar}{rgb}{.8,0,0}
\definecolor{greenmar}{rgb}{0,.5,0}

\newcommand{\ice}[1]{\relax}
\ChTitleVar{\Large\rm\bf}
\ChNameVar{\Large\rm\bf}

\begin{document}

\pagestyle{plain}

\setcounter{tocdepth}{1}
\pagenumbering{arabic}
\setlength{\fboxrule}{0.5 mm} 

\begin{flushright}
TTP12-012\\
SFB/CPP-12-23
\par\end{flushright}

\begin{center}
{\Huge Three-loop $\beta$-functions for \mbox{top-Yukawa} and the Higgs self-interaction 
in the Standard Model}
\vskip 0.3cm
K. G. Chetyrkin$^{\;a}$, M. F. Zoller$^{\;a}$\\[1ex]
{\small $^{a}$ Institut f\"ur Theoretische Teilchenphysik, Karlsruhe
  Institute of Technology (KIT), D-76128 Karlsruhe, Germany}
\vskip 0.5cm
{\bf Abstract}\\[1ex]
\end{center}
We analytically compute the dominant contributions to the
$\beta$-functions for the top-Yukawa coupling, the strong coupling and
the Higgs self-coupling as well as the anomalous dimensions of the scalar, gluon and quark fields 
in the unbroken phase of the Standard Model at three-loop level.
These are mainly the QCD and top-Yukawa corrections. The contributions
from the Higgs self-interaction which are negligible for the running
of the top-Yukawa and the strong coupling but important for the
running of the Higgs self-coupling are also evaluated.

\section{Introduction}
Using perturbation theory in any renormalizable quantum field theory
comes with the price that the parameters of this theory, e.g. the
couplings and masses, in general depend on the renormalization scale
$\mu$. 
The precise description of  the evolution  of these parameters with the energy
scale  is  an important task in any
model. This is done by means of the Renormalization Group functions, that is the
$\beta$-functions and anomalous dimensions.
The knowledge of the three-loop contributions to the $\beta$-functions for the
Standard Model (SM) and its extensions is important for physics at the very high energy 
frontier and for cosmology. Here are some examples: The running
of the gauge couplings plays an important role for the construction of Grand
Unified theories of the strong and electroweak interactions. The $\beta$-functions
for the scalar self-interaction and for the top quark Yukawa coupling constant 
are important for the analysis of Higgs-inflation in the SM \cite{Bezrukov:2007ep,Bezrukov:2008ej,DeSimone:2008ei,Bezrukov:2009db}. 
The current investigations of these issues are based on the
two-loop approximation. The inclusion of the next order could be essential.
Further, in a recent work \cite{Shaposhnikov:2009pv},
the possibility has been discussed that the SM, supplemented by the asymptotically safe gravity could play the role
of a fundamental, rather than effective field theory. Within this framework
the mass of the Higgs boson has been predicted to be approximately \mbox{126 GeV}.
The theoretical uncertainty of the prediction is about \mbox{2 GeV}. 

Recent exciting evidence  from  several SM-like Higgs
search channels at both the CERN Large Hadron Collider
and the Fermilab Tevatron \cite{ATLAS:2012ae,Chatrchyan:2012tx,TEVNPH:2012ab}
point to the possibility of a SM Higgs boson with a mass in the vicinity of 125 GeV which is in truly
remarkable agreement with the aforementioned prediction.\footnote{Note that  the boundary condition
$\lambda(M_{Planck})=0$, leading to the prediction of the Higgs mass close the 
the experimental evidence, has been also discussed recently in \cite{Holthausen:2011aa}.} 
This calls for more precise calculations, in particular of $\beta$-functions, in the SM.
In the present  paper we are particularly interested in the evolution of the Higgs self-coupling
as well  as the top-Yukawa coupling in the SM.

The underlying
gauge group of the SM is an $\text{SU}_{\sss{C}}(3)\times \text{SU}(2)\times
\text{U}_{\sss{Y}}(1)$ which is spontaneously broken to
$\text{SU}_{\sss{C}}(3)\times \text{U}_{\sss{Q}}(1)$ at the electroweak scale. As
the renormalization constants for fields and vertices do not depend on
masses and external momenta in the $\overline{\text{MS}}$-scheme, we
will perform   our calculations in the unbroken phase of the SM. 

The most important contributions to the running
of the Higgs self-coupling $\lambda$ arise from the top-Yukawa coupling
and the strong sector.  All other Yukawa couplings are significantly
smaller due to to the smallness of the respective quark masses. From the top
mass \mbox{$M_t \approx 172.9$ GeV} we get the Yukawa coupling at this scale
\mbox{$\yt(M_t)=\sqrt{2}\f{M_t}{v}\approx 1$} where \mbox{$v\approx246.2$ GeV} is
proportional to the vacuum expectation value of the scalar field
$\Phi$ from which results the Higgs field after the spontaneous
symmetry breaking:
$|\langle \Phi \rangle|=\f{v}{\sqrt{2}} $.

The next Yukawa coupling to be considered would be
\mbox{$y_b=\sqrt{2}\f{M_b}{v}\approx 0.02$}. The strong coupling at the scale
of the Z boson mass is \mbox{$\gs(M_Z)\approx 1.22$} whereas the electroweak
couplings 
\mbox{$g_1(M_Z)=\f{\sqrt{4\pi\alpha}}{\cos{\theta_W}} \approx
0.36$} and \mbox{$g_2(M_Z)=\f{\sqrt{4\pi\alpha}}{\sin{\theta_W}} \approx
0.65$} 
give much smaller contributions which are further suppressed by
the isospin and hypercharge factors. For this reason we will consider
a simplified version of the SM or - as one could also see
it - a minimal extension of QCD by setting \mbox{$g_1 =g_2=0$} in our calculation.
For a Higgs mass of \mbox{$125$ GeV} the value of the Higgs self-interaction 
would be $\lambda(M_H)\approx 0.13$ at the scale of the Higgs mass. The relevance of this parameter will be examined in
section \ref{setup}.

The outline of the work is as follows. In the next section we discuss the
main definitions and the general setup of our work. Section \ref{calc}
deals with the technical details, including the treatment of
$\gamma_5$.  In sections \ref{res:beta} and \ref{res:AD} we present
our results for the $\beta$-functions of the top-Yukawa, the strong and
the Higgs self-couplings and the relevant field anomalous dimensions. 
The numerical influence of the computed three-loop
corrections  on the evolution of the quartic Higgs coupling is discussed in section
\ref{la:evolution}. For this analysis we will include the already known
contributions with $g_1$ and $g_2$  at one-loop and two-loop level. Finally, section \ref{last}
contains our conclusions and acknowledgements.

All our results for $\beta$-functions and anomalous dimensions can  be retrieved from
\\
\texttt{\bf http://www-ttp.particle.uni-karlsruhe.de/Progdata/ttp12/ttp12-012/}

\section{General Setup \label{setup} }

The Lagrangian of our model consists of three pieces:
\be
\ssL=\ssL_{\sss{QCD}}+\ssL_{\sss{\yt}}+\ssL_{\sss{\Phi}}.
\ee
The QCD part is defined by
\be
\begin{split}
\ssL_{\sss{QCD}}=&-\f{1}{4}G^a_{\mu \nu} G^{a\,\mu \nu}-\f{1}{2 (1-\xi)}\lb\p_\mu A^{a\,\mu}\rb^2 
+\p_\mu \bar{c}^a \p^{\mu}c^a+\gs f^{abc}\,\p_\mu \bar{c}^a A^{b\,\mu} c^c \\
&+\sum\limits_q\left\{\f{i}{2}\bar{q}\overleftrightarrow{\slashed{\p}}q+ \gs \bar{q}\slashed{A}^a T^a q\right\}
\end{split} 
\label{LQCD} 
{},
\ee
where $q$ runs over all quark flavours, the gluon field strength tensor is given by
\be
G^a_{\mu \nu}=\p_\mu A^a_\nu - \p_\nu A^a_\mu + \gs f^{abc}A^b_\mu A^c_\nu
\ee
and $f^{abc}$ are the structure constants of the colour gauge   group with the generators $T^a$:
\be \left[ T^a,T^b \right]=if^{abc}T^c.\ee
The complex scalar field $\Phi$ and the left-handed parts of the top and bottom quarks $t_{\sss{L}}$ and $b_{\sss{L}}$ are doublets 
under SU$(2)$:
\be
\Phi=\vv{\Phi_1}{\Phi_2}, \qquad Q_{\sss{L}}=\vv{t}{b}_{\sss{L}}
{}.
\ee
Setting all Yukawa couplings to zero except for the top coupling $\yt$ the Lagrangian for the Yukawa sector is given by 
\be
\begin{split}
\ssL_{\sss{\yt}}&=-\yt \left\{\bar{t}_{\sss{R}}\Phi^{\dagger\,c}Q_{\sss{L}}+\bar{Q}_{\sss{L}}\Phi^c t_{\sss{R}}\right\}\\
&=-\yt \left\{ 
\bar{t}_{\sss{R}}\lb \Phi_2, -\Phi_1\rb\cd \vv{t}{b}_{\sss{L}}
+\,\lb \bar{t}, \bar{b}\rb_{\sss{L}}\cd \vv{\Phi^{*}_2}{-\Phi^{*}_1}\,t_{\sss{R}}
\right\}\\
&=-\yt \left\{ \lb \bar{t} P_{\sss{R}} t\rb \Phi^{*}_2+\lb\bar{t} P_{\sss{L}} t\rb \Phi_2
-\lb\bar{b} P_{\sss{R}} t\rb \Phi^{*}_1-\lb \bar{t} P_{\sss{L}} b\rb \Phi_1
\right\}.
\end{split} 
\label{LYuk} 
\ee
Finally, we have the scalar sector of the model
\be
\begin{split}
\ssL_{\sss{\Phi}}=\p_\mu \Phi^\dagger \p^\mu \Phi-m^2\Phi^\dagger\Phi-\lambda \lb\Phi^\dagger\Phi\rb^2.
\end{split} 
\label{LPhi} 
\ee  
The indices L and R indicate the left- and right-handed part of the fields as obtained by the projectors
\be 
P_{\sss{L}}=\f{1}{2}\lb 1-\gamma_5\rb \qquad P_{\sss{R}}=\f{1}{2}\lb 1+\gamma_5\rb
{}.
\ee

This model is renormalized with the counterterm Lagrangians
\be
\begin{split}
\delta\!\ssL_{\sss{QCD}}=&-\f{1}{4}\delta\! Z^{(2g)}_3 \lb \p_\mu A^a_\nu - \p_\nu A^a_\mu \rb^2
-\f{1}{2}\delta\! Z^{(3g)}_1 \gs f^{abc}\lb \p_\mu A^a_\nu - \p_\nu A^a_\mu \rb A^b_\mu A^c_\nu\\
&-\f{1}{4}\delta\! Z^{(4g)}_1 \gs^2 \lb f^{abc} A^b_\mu A^c_\nu \rb^2
+\delta\! Z^{(2c)}_3 \p_\mu \bar{c}^a \p^{\mu}c^a+\delta\! Z^{(ccg)}_1 \gs f^{abc}\,\p_\mu \bar{c}^a A^{b\,\mu} c^c \\
&+\sum\limits_q\left\{
\f{i}{2}\bar{q}\overleftrightarrow{\slashed{\p}}\left[\delta\! Z^{(2q)}_{2,L}P_{\sss{L}}+\delta\! Z^{(2q)}_{2,R}P_{\sss{R}}\right]q
+\gs \bar{q}\slashed{A}^a T^a\left[\delta\! Z^{(qqg)}_{1,L}P_{\sss{L}} +\delta\! Z^{(qqg)}_{1,R}P_{\sss{R}}\right]q
\right\}
\end{split} 
\label{LQCDc} 
\ee
for the QCD part,
\be
\delta\!\ssL_{\sss{Yukawa}}=-\delta\! Z^{(tb\Phi)}_1 
\yt\left\{ \lb \bar{t} P_{\sss{R}} t\rb \Phi^{*}_2+\lb\bar{t} P_{\sss{L}} t\rb \Phi_2
-\lb\bar{b} P_{\sss{R}} t\rb \Phi^{*}_1-\lb \bar{t} P_{\sss{L}} b\rb \Phi_1
\right\}
\label{LYukc} 
\ee
for the Yukawa part and
\be
\begin{split}
\delta\!\ssL_{\sss{\Phi}}=\delta\! Z_2^{(2\Phi)}\p_\mu \Phi^\dagger \p^\mu \Phi-m^2\, \delta\! Z_{\Phi^2}\Phi^\dagger\Phi
+\delta\! Z_1^{(4\Phi)}\lb\Phi^\dagger\Phi\rb^2
\end{split} 
\label{LPhic} 
\ee  
for the $\Phi$-sector. 
Note that in general the left- and right-handed parts of quark fields and quark-vertices are renormalized
differently.\footnote{In our case this is true for the quark fields participating in the Yukawa sector.
As we do not consider the electroweak interaction here and neglect all Yukawa couplings except for $\yt$ the light quark
fields u,d,s and c have the same renormalization constant for the left- and right-handed part.}  
So the renormalization constant for the strong gauge coupling $\gs$ can be obtained for example from
\be 
Z_{\gs}=\f{Z^{(ttg)}_{1,L}}{Z^{(2t)}_{2,L}\sqrt{Z^{(2g)}_3}}=
\f{Z^{(ttg)}_{1,R}}{Z^{(2t)}_{2,R}\sqrt{Z^{(2g)}_3}} \ee 
or the renormalization constant for $\yt$ from
\be 
Z_{\yt}=\f{Z^{(tb\Phi)}_1}{\sqrt{Z^{(2t)}_{2,L}Z^{(2t)}_{2,R}Z_2^{(2\Phi)}}}
{}. 
\ee
Here the renormalization constants have been defined in a minimal way as
\be Z=1+\delta 
Z
{}, 
\ee
with $\delta Z$ containing only poles in the regulating parameter  $\eps = (4-D)/2$ of the dimensional regularization and
$D$ being the engineering space-time dimension.
The Higgs self-coupling $\lambda$ is related to the Higgs mass at tree level $M_H$ via 
\be 
\lambda=\f{M_H^2}{2v^2} \label{lambdatree} 
{},
\ee
which for $M_H=125$ GeV yields $\lambda(M_H=125 \text{GeV}) \approx 0.13$.
For the running of the top-Yukawa coupling the contribution from
$\lambda$ is negligible compared to the top-Yukawa and strong coupling. The corresponding four-$\Phi$ vertex is nevertheless needed
for the renormalization at three-loop level, namely to kill the subdivergence from the fermion loop in diagrams like Fig. \ref{diasZphtb} (d).
The $\beta$-function for a coupling X is defined as
\be
\beta_{\sss{X}}=\mu^2\f{d X}{d \mu^2}    =\sum \limits_{n=1}^{\infty} \f{1}{(16\pi^2)^{n}}\,\beta_{\sss{X}}^{(n)}
\ee
and is given as a power series in all couplings of the model, namely $\gs$, $\yt$ and $\lambda$. 
Note that the $\beta$-functions $\beta_{\sss{\gs}}$ and $\beta_{\sss{\yt}}$ are proportional to $\gs$ and $\yt$ respectively whereas
$\beta_{\sss{\lambda}}$ has one part proportional to $\lambda$ and one part proportional to $\yt^4$ with no $\lambda$-dependence at all.
The anomalous dimension of a field $f$ is defined as
\be
\gamma^{\sss{f}}_2=-\mu^2\f{d \text{ln} Z_{\sss{f}}^{-1}}{d \mu^2}    =
\sum \limits_{n=1}^{\infty} \f{1}{(16\pi^2)^{n}}\,\gamma_2^{{\sss{f}}\,(n)}
{},
\ee
where $Z_{\sss{f}}$ is the field strength renormalization constant for the respective field.\footnote{For an $n$-point vertex $V$ or a 
mass $m$ the anomalous dimension is defined as \mbox{$\gamma^{\sss{V}}_{n}=-\mu^2\f{d \text{ln} Z_{\sss{V}}}{d \mu^2}$} or
\mbox{$\gamma_{\sss{m}}=-\mu^2\f{d \text{ln} Z_{\sss{m}}}{d \mu^2}$}. For a field we take the inverse renormalization constant.}
The $\beta$-functions for all couplings are independent of the gauge parameter $\xi$ whereas the anomalous dimensions of the fields are
not.

\section{Calculation \label{calc} }
As we are interested in $\beta$-functions it is enough to compute the
UV divergent part of all diagrams in order to determine the necessary
renormalization constants. In the $\overline{\text{MS}}$-scheme the latter depends only polynomially on external momenta and masses.
Therefore most of our renormalization constants could be computed from massless
propagator-like diagrams using the FORM 3 \cite{Vermaseren:2000nd} version\footnote{ 
The program can be downloaded  from \\
{\bf \hspace*{2cm} \texttt{http://www.nikhef.nl/\~\,form/maindir/packages/mincer/mincer.html}}}
of the  package MINCER \cite{MINCER}.
In some diagrams, e.g. with four external $\Phi$-fields where two
external momenta are set to zero, this leads to IR divergences which
mix with the UV ones in dimensional regularization. Another convenient
method to compute renormalization constants has been suggested in
\cite{Misiak:1994zw} and elaborated in the context of three-loop
calculations in \cite{beta_den_comp}.  The idea is an exact decomposition of all
propagators using an auxiliary mass parameter $M^2$:
\be
\f{1}{(q+p)^2}=\f{1}{q^2-M^2}+\f{-p^2-2q\cd p-M^2}{q^2-M^2}\f{1}{(q+p)^2}
{},
\ee
where $p$ is a combination of external and $q$ of internal momenta. This can be done recursively until the power in the denominator
of the last term is high enough for this contribution to be finite, e.g.
\be 
\begin{split}
\f{1}{(q+p)^2}=& \f{1}{q^2-M^2}
+
\f{-p^2-2q\cd p}{(q^2-M^2)^2}
+\f{(-p^2-2q\cd p)^2}{(q^2-M^2)^3}-\f{M^2}{(q^2-M^2)^2}\\
&+\f{M^2(M^2+2p^2+4q \cd p)}{(q^2-M^2)^3}   +
\f{(-p^2-2q\cd p-M^2)^3}{(q^2-M^2)^3}\f{1}{(q+p)^2}.
\end{split}
\ee
As the result is independent of $M^2$ we can omit the contributions $\sim M^2$ in the above decomposition
as long as we introduce counterterms into the Lagrangian to cancel $M^2$-dependent subdivergences. In our case only
a term 
$\f{M^2}{2}\delta\!Z_{\sss{M^2}}^{(2g)}\,A_\mu^a A^{a\,\mu}$ 
and 
$\f{M^2}{2}\delta\!Z_{\sss{M^2}}^{(2\Phi)}\, \Phi^\dagger\Phi$
are possible.\footnote{
Counterterms $\sim M$ that would arise for fermions cannot appear because we have no $M$ in the numerators
of propagators. The ghost mass term $\f{M^2}{2}\delta\!Z_{\sss{M^2}}^{(2c)}\,\bar{c}^a c^a$ does not appear because of the momentum dependence
of the ghost-gluon-vertex.} 
The first one is not gauge invariant but this does not matter as it is only used
for the cancellation of subdivergences which works nevertheless. 
This method effectively amounts to introducing the same auxiliary mass
parameter $M^2$ in every denominator of propagators and all possible
$M^2$-counterterms. We expand in the external momenta\footnote{This
method also works in massive theories. In this case we expand in the
physical masses as well.} and arrive at massive tadpole diagrams with
one scale $M$.   Due to the auxiliary mass no IR divergences can appear 
while the  UV counterterms which we are interested in  (that is the ones without any 
dependence on the  auxiliary mass M) will stay untouched.

For the calculation of massive tadpoles we have used the FORM-based program
MATAD \cite{MATAD}. Where possible, i.e. for the propagators
and three point functions, we have  employed  both  the MINCER and the MATAD setups 
which served as an extra check. To generate the
diagrams we used QGRAF \cite{QGRAF} and to compute the 
colour factors the FORM package COLOR \cite{COLOR}.

An important aspect of calculations such as these is the proper treatment of $\gamma_5$. As is well known, a naive treatment of $\gamma_5$
can be applied if it only appears in an external fermion line. In fermion loops we have to be more careful. In four dimensions we define
\be \gamma_5=i\gamma^0\gamma^1\gamma^2\gamma^3=\f{i}{4!}\eps_{\mu\nu\rho\sigma} \gamma^\mu\gamma^\nu\gamma^\rho\gamma^\sigma \text{ with }
 \eps_{0123}=1=-\eps^{0123} \label{gamma5} 
{}.
\ee
In Fig. 1 to 5 we show a few diagrams that had to be calculated for the various ingredients of our final result.
In order to have a contribution from a fermion loop with one $\gamma_5$ in it at least four free Lorentz indices or momenta
on the external lines of the minimal subgraph containing this fermion loop are required.
{These can be indices
from the gluon vertices or the internal momenta from other loops which act as external momenta to 
the minimal subgraph containing the fermion loop in question.
External momenta of the whole diagram can be set to zero as the renormalization constants in the $\overline{\text{MS}}$-scheme do not
depend on those.

Consider for example one of the fermion loops in 
Fig.\ref{diasZ4ph} (c). The momenta on the two external $\Phi$-legs can be set to zero. Then we have two indices from the
gluon lines attached to our fermion loop and one loop momentum going through the two gluons and acting as an external momentum
to the subgraph containing only our fermion loop. This is not enough to have a non-naive $\gamma_5$ contribution from this graph.}

\begin{figure}[h!]
  \begin{tabular}{lll}
    \begin{picture}(140,100) (0,0)
    \SetWidth{0.5}
    \SetColor{Black}
    \DashArrowLine(70,95)(70,75){4}
    \ArrowLine(70,75)(110,30)
    \ArrowLine(110,30)(30,30)
    \ArrowLine(30,30)(70,75)
    \DashArrowLine(30,30)(30,0){4}
    \DashArrowLine(110,30)(110,0){4}
    \ArrowLine(5,0)(30,0)
    \ArrowLine(30,0)(110,0)
    \ArrowLine(110,0)(135,0)
    \Text(10,-15)[lb]{\Large{\Black{$t$}}}
    \Text(125,-15)[lb]{\Large{\Black{$t$}}}
    \Text(75,90)[lb]{\Large{\Black{$\Phi$}}}
    \Text(5,75)[lb]{\Large{\Black{(a)}}}
  \end{picture}
 &     \begin{picture}(140,100) (0,0)
    \SetWidth{0.5}
    \SetColor{Black}
    \DashArrowLine(70,95)(70,75){4}
    \ArrowLine(70,75)(110,30)
    \ArrowLine(110,30)(30,30)
    \ArrowLine(30,30)(70,75)
    \Gluon(30,30)(30,0){4}{4}
    \DashArrowLine(110,30)(110,0){4}
    \ArrowLine(5,0)(30,0)
    \ArrowLine(30,0)(110,0)
    \ArrowLine(110,0)(135,0)
    \Text(10,-15)[lb]{\Large{\Black{$t$}}}
    \Text(125,-15)[lb]{\Large{\Black{$t$}}}
    \Text(75,90)[lb]{\Large{\Black{$\Phi$}}}
    \Text(5,75)[lb]{\Large{\Black{(b)}}}
  \end{picture}
 &     \begin{picture}(140,100) (0,0)
    \SetWidth{0.5}
    \SetColor{Black}
    \DashArrowLine(70,95)(70,75){4}
    \ArrowLine(70,75)(110,30)
    \ArrowLine(110,30)(30,30)
    \ArrowLine(30,30)(70,75)
    \Gluon(30,30)(30,0){4}{4}
    \Gluon(110,30)(110,0){4}{4}
    \ArrowLine(5,0)(30,0)
    \ArrowLine(30,0)(110,0)
    \ArrowLine(110,0)(135,0)
    \Text(10,-15)[lb]{\Large{\Black{$t$}}}
    \Text(125,-15)[lb]{\Large{\Black{$t$}}}
    \Text(75,90)[lb]{\Large{\Black{$\Phi$}}}
    \Text(5,75)[lb]{\Large{\Black{(c)}}}
  \end{picture}\\[5ex]
    \begin{picture}(140,100) (0,0)
    \SetWidth{0.5}
    \SetColor{Black}
    \DashArrowLine(70,95)(70,75){4}
    \ArrowLine(70,75)(110,30)
    \ArrowLine(110,30)(70,30)
    \ArrowLine(70,30)(30,30)
    \ArrowLine(30,30)(70,75)
    \DashArrowLine(30,30)(30,0){4}
    \DashArrowLine(70,0)(70,30){4}
    \DashArrowLine(110,30)(110,0){4}
    \ArrowLine(5,0)(30,0)
    \ArrowLine(30,0)(70,0)
    \ArrowLine(70,0)(110,0)
    \ArrowLine(110,0)(135,0)
    \Text(10,-15)[lb]{\Large{\Black{$t$}}}
    \Text(125,-15)[lb]{\Large{\Black{$t$}}}
    \Text(75,90)[lb]{\Large{\Black{$\Phi$}}}
    \Text(5,75)[lb]{\Large{\Black{(d)}}}
  \end{picture}
&    \begin{picture}(140,100) (0,0)
    \SetWidth{0.5}
    \SetColor{Black}
    \DashArrowLine(70,95)(70,75){4}
    \ArrowLine(70,75)(110,30)
    \ArrowLine(110,30)(70,30)
    \ArrowLine(70,30)(30,30)
    \ArrowLine(30,30)(70,75)
    \Gluon(30,30)(30,0){4}{4}
    \Gluon(70,30)(70,0){4}{4}
    \Gluon(110,30)(110,0){4}{4}
    \ArrowLine(5,0)(30,0)
    \ArrowLine(30,0)(70,0)
    \ArrowLine(70,0)(110,0)
    \ArrowLine(110,0)(135,0)
    \Text(10,-15)[lb]{\Large{\Black{$t$}}}
    \Text(125,-15)[lb]{\Large{\Black{$t$}}}
    \Text(75,90)[lb]{\Large{\Black{$\Phi$}}}
    \Text(5,75)[lb]{\Large{\Black{(e)}}}
  \end{picture}
&    \begin{picture}(140,100) (0,0)
    \SetWidth{0.5}
    \SetColor{Black}
    \DashArrowLine(70,95)(70,75){4}
    \ArrowLine(70,75)(110,30)
    \ArrowLine(110,30)(70,30)
    \ArrowLine(70,30)(30,30)
    \ArrowLine(30,30)(70,75)
    \Gluon(30,30)(30,0){4}{4}
    \Gluon(70,30)(70,0){4}{4}
    \DashArrowLine(110,30)(110,0){4}
    \ArrowLine(5,0)(30,0)
    \ArrowLine(30,0)(70,0)
    \ArrowLine(70,0)(110,0)
    \ArrowLine(110,0)(135,0)
    \Text(10,-15)[lb]{\Large{\Black{$t$}}}
    \Text(125,-15)[lb]{\Large{\Black{$t$}}}
    \Text(75,90)[lb]{\Large{\Black{$\Phi$}}}
    \Text(5,75)[lb]{\Large{\Black{(f)}}}
  \end{picture} \\[4ex]
\end{tabular}
\caption{Some diagrams contributing to $Z^{(tb\Phi)}_1$}
\label{diasZphtb}
\end{figure}
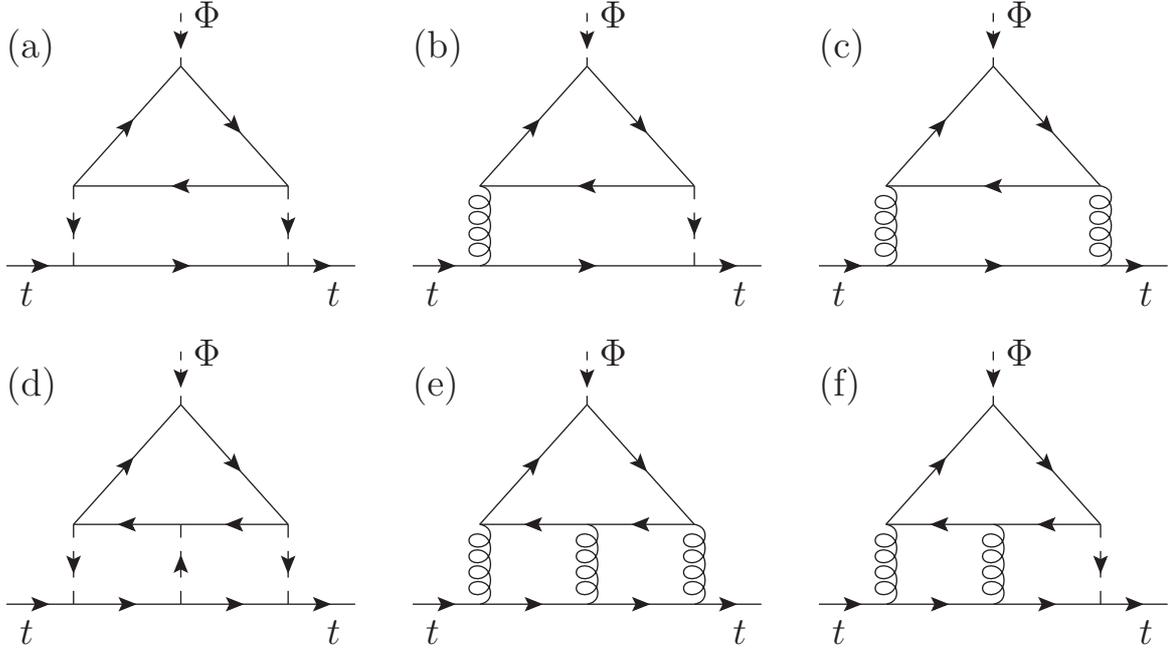

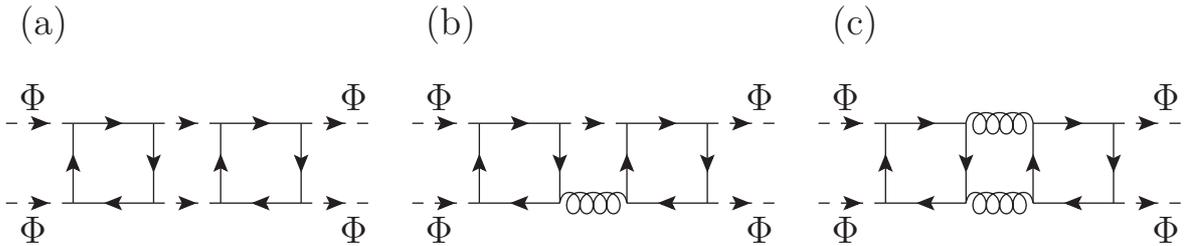
\begin{figure}[h!]
  \begin{tabular}{lll}
    \begin{picture}(140,100) (0,0)
    \SetWidth{0.5}
    \SetColor{Black}
    \DashArrowLine(0,15)(25,15){4}
    \DashArrowLine(0,45)(25,45){4}
    \DashArrowLine(110,15)(135,15){4}
    \DashArrowLine(110,45)(135,45){4}
    \ArrowLine(25,45)(55,45)
    \ArrowLine(55,15)(25,15)
    \ArrowLine(25,15)(25,45)
    \ArrowLine(55,45)(55,15)
    \DashArrowLine(55,15)(80,15){4}
    \DashArrowLine(55,45)(80,45){4}
    \ArrowLine(80,45)(110,45)
    \ArrowLine(110,15)(80,15)
    \ArrowLine(80,15)(80,45)
    \ArrowLine(110,45)(110,15)
    \Text(5,0)[lb]{\Large{\Black{$\Phi$}}}
    \Text(125,0)[lb]{\Large{\Black{$\Phi$}}}
    \Text(5,50)[lb]{\Large{\Black{$\Phi$}}}
    \Text(125,50)[lb]{\Large{\Black{$\Phi$}}}
    \Text(5,75)[lb]{\Large{\Black{(a)}}}
  \end{picture}
&    \begin{picture}(140,100) (0,0)
    \SetWidth{0.5}
    \SetColor{Black}
    \DashArrowLine(0,15)(25,15){4}
    \DashArrowLine(0,45)(25,45){4}
    \DashArrowLine(110,15)(135,15){4}
    \DashArrowLine(110,45)(135,45){4}
    \ArrowLine(25,45)(55,45)
    \ArrowLine(55,15)(25,15)
    \ArrowLine(25,15)(25,45)
    \ArrowLine(55,45)(55,15)
    \Gluon(55,15)(80,15){4}{4}
    \DashArrowLine(55,45)(80,45){4}
    \ArrowLine(80,45)(110,45)
    \ArrowLine(110,15)(80,15)
    \ArrowLine(80,15)(80,45)
    \ArrowLine(110,45)(110,15)
    \Text(5,0)[lb]{\Large{\Black{$\Phi$}}}
    \Text(125,0)[lb]{\Large{\Black{$\Phi$}}}
    \Text(5,50)[lb]{\Large{\Black{$\Phi$}}}
    \Text(125,50)[lb]{\Large{\Black{$\Phi$}}}
    \Text(5,75)[lb]{\Large{\Black{(b)}}}
  \end{picture}
&    \begin{picture}(140,100) (0,0)
    \SetWidth{0.5}
    \SetColor{Black}
    \DashArrowLine(0,15)(25,15){4}
    \DashArrowLine(0,45)(25,45){4}
    \DashArrowLine(110,15)(135,15){4}
    \DashArrowLine(110,45)(135,45){4}
    \ArrowLine(25,45)(55,45)
    \ArrowLine(55,15)(25,15)
    \ArrowLine(25,15)(25,45)
    \ArrowLine(55,45)(55,15)
    \Gluon(55,15)(80,15){4}{4}
    \Gluon(55,45)(80,45){4}{4}
    \ArrowLine(80,45)(110,45)
    \ArrowLine(110,15)(80,15)
    \ArrowLine(80,15)(80,45)
    \ArrowLine(110,45)(110,15)
    \Text(5,0)[lb]{\Large{\Black{$\Phi$}}}
    \Text(125,0)[lb]{\Large{\Black{$\Phi$}}}
    \Text(5,50)[lb]{\Large{\Black{$\Phi$}}}
    \Text(125,50)[lb]{\Large{\Black{$\Phi$}}}
    \Text(5,75)[lb]{\Large{\Black{(c)}}}
  \end{picture}
\end{tabular}
\caption{Some diagrams contributing to $Z^{(4\Phi)}_1$}
\label{diasZ4ph}
\end{figure}

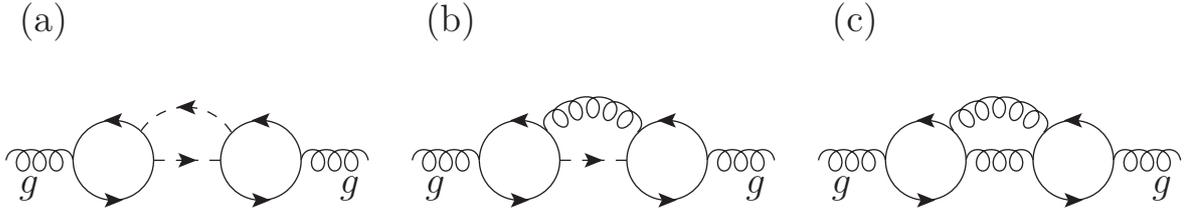
\begin{figure}[h!]
  \begin{tabular}{lll}
    \begin{picture}(140,100) (0,0)
    \SetWidth{0.5}
    \SetColor{Black}
    \Gluon(0,30)(25,30){4}{3}
    \Gluon(110,30)(135,30){4}{3}
    \ArrowArc(40,30)(15,180,0)
    \ArrowArc(40,30)(15,0,180)
    \ArrowArc(95,30)(15,180,0)
    \ArrowArc(95,30)(15,0,180)
    \DashArrowLine(55,30)(80,30){4}
    \DashArrowArc(67,30)(20,32,146){4}
    \Text(5,15)[lb]{\Large{\Black{$g$}}}
    \Text(125,15)[lb]{\Large{\Black{$g$}}}
    \Text(5,75)[lb]{\Large{\Black{(a)}}}
  \end{picture}
 &   \begin{picture}(140,100) (0,0)
    \SetWidth{0.5}
    \SetColor{Black}
    \Gluon(0,30)(25,30){4}{3}
    \Gluon(110,30)(135,30){4}{3}
    \ArrowArc(40,30)(15,180,0)
    \ArrowArc(40,30)(15,0,180)
    \ArrowArc(95,30)(15,180,0)
    \ArrowArc(95,30)(15,0,180)
    \DashArrowLine(55,30)(80,30){4}
    \GlueArc(67,30)(20,32,146){4}{5}
    \Text(5,15)[lb]{\Large{\Black{$g$}}}
    \Text(125,15)[lb]{\Large{\Black{$g$}}}
    \Text(5,75)[lb]{\Large{\Black{(b)}}}
  \end{picture}
&    \begin{picture}(140,100) (0,0)
    \SetWidth{0.5}
    \SetColor{Black}
    \Gluon(0,30)(25,30){4}{3}
    \Gluon(110,30)(135,30){4}{3}
    \ArrowArc(40,30)(15,180,0)
    \ArrowArc(40,30)(15,0,180)
    \ArrowArc(95,30)(15,180,0)
    \ArrowArc(95,30)(15,0,180)
    \Gluon(55,30)(80,30){4}{3}
    \GlueArc(67,30)(20,32,146){4}{5}
    \Text(5,15)[lb]{\Large{\Black{$g$}}}
    \Text(125,15)[lb]{\Large{\Black{$g$}}}
    \Text(5,75)[lb]{\Large{\Black{(c)}}}
  \end{picture}
\end{tabular}
\caption{Some diagrams contributing to $Z^{(2g)}_3$}
\label{diasgprop}
\end{figure}

For this reason diagrams like Fig.\ref{diasZphtb} (a,b,c,d),
Fig.\ref{diasZ4ph} (a,b,c), Fig.\ref{diasgprop} (a,b), Fig.\ref{diasphprop} (a,b,c), Fig.\ref{diastprop} (a,b) can be treated naively.
Fig.\ref{diasgprop} (c) has enough indices and momenta but no $\gamma_5$ in it.
Diagrams like Fig.\ref{diasZphtb} (c,e), Fig.\ref{diasZ4ph} (b), Fig.\ref{diasgprop} (b), Fig.\ref{diasphprop} (a,c),
Fig.\ref{diastprop} (a,c) are zero because of an odd number of $\gamma$-matrices in at least one fermion loop.
And diagrams like Fig.\ref{diasZphtb} (b), Fig.\ref{diasZ4ph} (b), Fig.\ref{diasgprop} (a), Fig.\ref{diasphprop} (b),
Fig.\ref{diastprop} (b) are zero because of their colour structure. The only problematic type is Fig.\ref{diasZphtb} (f) which fortunately only contributes a $\f{1}{\eps}$
pole and can therefore be treated as described in \cite{'tHooft:1972fi}. We use the fact that $\gamma_5$ anticommutes with every other
$\gamma$-matrix in four dimensions and that $\gamma_5^2=1$. Then we apply relation \eqref{gamma5} for the case when one $\gamma_5$
remains on each fermion line. The two $\eps_{\mu\nu\rho\sigma}$ can be rewritten as a combination of metric tensors which can be handled
in dimensional regularization. The error we make with this treatment is of order $\eps$ and does therefore not affect the pole part of our
result.

\begin{figure}[h!]
  \begin{tabular}{lll}
    \begin{picture}(140,100) (0,0)
    \SetWidth{0.5}
    \SetColor{Black}
    \DashArrowLine(0,30)(25,30){4}
    \DashArrowLine(110,30)(135,30){4}
    \ArrowArc(40,30)(15,180,0)
    \ArrowArc(40,30)(15,0,180)
    \ArrowArc(95,30)(15,180,0)
    \ArrowArc(95,30)(15,0,180)
    \DashArrowLine(55,30)(80,30){4}
    \DashArrowArc(67,30)(20,32,146){4}
    \Text(5,15)[lb]{\Large{\Black{$\Phi$}}}
    \Text(125,15)[lb]{\Large{\Black{$\Phi$}}}
    \Text(5,75)[lb]{\Large{\Black{(a)}}}
  \end{picture}
 &   \begin{picture}(140,100) (0,0)
    \SetWidth{0.5}
    \SetColor{Black}
    \DashArrowLine(0,30)(25,30){4}
    \DashArrowLine(110,30)(135,30){4}
    \ArrowArc(40,30)(15,180,0)
    \ArrowArc(40,30)(15,0,180)
    \ArrowArc(95,30)(15,180,0)
    \ArrowArc(95,30)(15,0,180)
    \DashArrowLine(55,30)(80,30){4}
    \GlueArc(67,30)(20,32,146){4}{5}
    \Text(5,15)[lb]{\Large{\Black{$\Phi$}}}
    \Text(125,15)[lb]{\Large{\Black{$\Phi$}}}
    \Text(5,75)[lb]{\Large{\Black{(b)}}}
  \end{picture}
&    \begin{picture}(140,100) (0,0)
    \SetWidth{0.5}
    \SetColor{Black}
    \DashArrowLine(0,30)(25,30){4}
    \DashArrowLine(110,30)(135,30){4}
    \ArrowArc(40,30)(15,180,0)
    \ArrowArc(40,30)(15,0,180)
    \ArrowArc(95,30)(15,180,0)
    \ArrowArc(95,30)(15,0,180)
    \Gluon(55,30)(80,30){4}{3}
    \GlueArc(67,30)(20,32,146){4}{5}
    \Text(5,15)[lb]{\Large{\Black{$\Phi$}}}
    \Text(125,15)[lb]{\Large{\Black{$\Phi$}}}
    \Text(5,75)[lb]{\Large{\Black{(c)}}}
  \end{picture}
\end{tabular}
\caption{Some diagrams contributing to $Z^{(2\Phi)}_2$}
\label{diasphprop}
\end{figure}
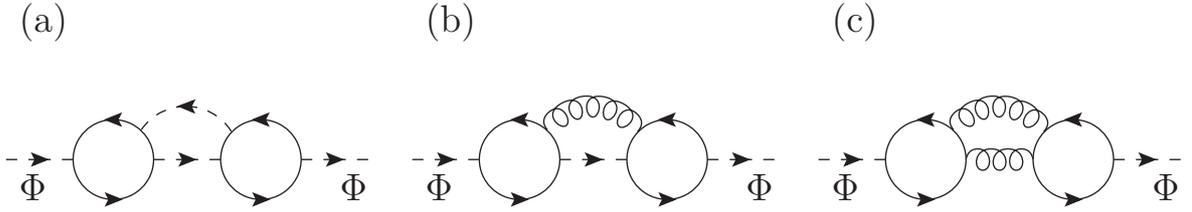

\begin{figure}[h!]
  \begin{tabular}{lll}
    \begin{picture}(140,100) (0,0)
    \SetWidth{0.5}
    \SetColor{Black}
    \ArrowArcn(70,30)(40,180,0)
    \ArrowLine(110,30)(70,30)
    \ArrowLine(70,30)(30,30)
    \DashArrowLine(30,30)(30,0){4}
    \DashArrowLine(70,0)(70,30){4}
    \DashArrowLine(110,30)(110,0){4}
    \ArrowLine(5,0)(30,0)
    \ArrowLine(30,0)(70,0)
    \ArrowLine(70,0)(110,0)
    \ArrowLine(110,0)(135,0)
    \Text(10,-15)[lb]{\Large{\Black{$t$}}}
    \Text(125,-15)[lb]{\Large{\Black{$t$}}}
    \Text(5,75)[lb]{\Large{\Black{(a)}}}
  \end{picture}
&    \begin{picture}(140,100) (0,0)
    \SetWidth{0.5}
    \SetColor{Black}
    \ArrowArcn(70,30)(40,180,0)
    \ArrowLine(110,30)(70,30)
    \ArrowLine(70,30)(30,30)
    \Gluon(30,30)(30,0){4}{4}
    \Gluon(70,30)(70,0){4}{4}
    \Gluon(110,30)(110,0){4}{4}
    \ArrowLine(5,0)(30,0)
    \ArrowLine(30,0)(70,0)
    \ArrowLine(70,0)(110,0)
    \ArrowLine(110,0)(135,0)
    \Text(10,-15)[lb]{\Large{\Black{$t$}}}
    \Text(125,-15)[lb]{\Large{\Black{$t$}}}
    \Text(5,75)[lb]{\Large{\Black{(b)}}}
  \end{picture}
&    \begin{picture}(140,100) (0,0)
    \SetWidth{0.5}
    \SetColor{Black}
    \ArrowArcn(70,30)(40,180,0)
    \ArrowLine(110,30)(70,30)
    \ArrowLine(70,30)(30,30)
    \Gluon(30,30)(30,0){4}{4}
    \Gluon(70,30)(70,0){4}{4}
    \DashArrowLine(110,30)(110,0){4}
    \ArrowLine(5,0)(30,0)
    \ArrowLine(30,0)(70,0)
    \ArrowLine(70,0)(110,0)
    \ArrowLine(110,0)(135,0)
    \Text(10,-15)[lb]{\Large{\Black{$t$}}}
    \Text(125,-15)[lb]{\Large{\Black{$t$}}}
    \Text(5,75)[lb]{\Large{\Black{(c)}}}
  \end{picture} \\[4ex]
\end{tabular}
\caption{Some diagrams contributing to $Z^{(2t)}_{2,L/R}$}
\label{diastprop}
\end{figure}
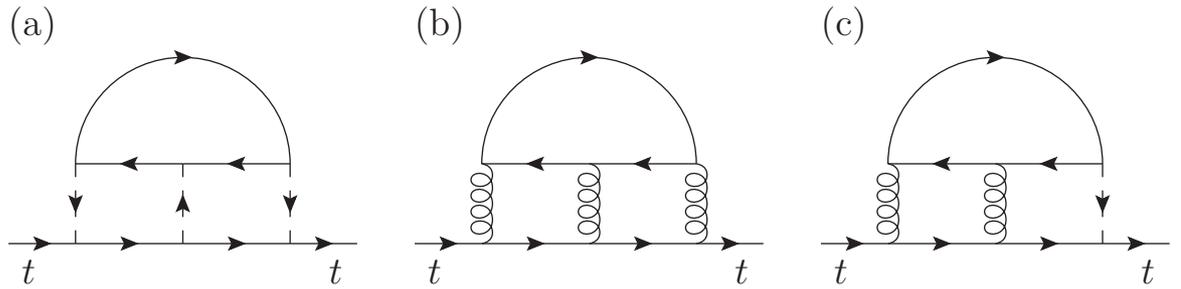

\section{Results for the $\beta$-functions \label{res:beta}}
First we give the results for the three-loop $\beta$-functions of 
couplings $\lambda$, $\yt$ and $\gs$ with the general gauge group
factors for the strong interacting sector. 
Below  $\cf$ and $\ca$ are the quadratic Casimir
operators of the 
quark  and the adjoint representation of the corresponding Lie algebra,
$\dR$ is the dimension of the quark representation,
$\tr$ is defined so that  \mbox{$\tr \delta^{ab}=\textbf{Tr}\lb T^a T^b\rb$}  
is the trace of two group generators of the quark representation.\footnote{For an SU$(N)$ gauge group these are $\dR=N$,
$\ca=2\tr N$ and $\cf=\tr\lb N-\f{1}{N}\rb$.}
For  QCD (colour gauge group SU$(3)$) we have $\cf =4/3\,,\, \ca=3\,,\,\tr=1/2$ and $\dR = 3$.
Furthermore we denote the number of fermions by $\Nf=\Nl+1$.
\be
\begin{split}
\beta_{\sss{\lambda}}^{(1)}=& 12\, \lambda^2              
       +  2 \dR\,\yt^2 \lambda            
       - \dR\,  \yt^4     ,\\
\beta_{\sss{\lambda}}^{(2)}=&      
       - 156\,\lambda^3                   
       - 24 \dR \,\yt^2 \lambda^2   
       - \f{1}{2} \dR \,\yt^4 \lambda     
       + 5 \dR \yt^6       \\
       &+ 10 \cf \dR\, \gs^2 \yt^2 \lambda  
         - 4 \cf \dR\, \gs^2 \yt^4   ,\\
\beta_{\sss{\lambda}}^{(3)}=& 
 \lambda^4   \left(           3588          + 2016 \zeta_{3}          \right)
       +291 \dR \, \yt^2 \lambda^3             
       + \yt^4 \lambda^2   \left( \f{789}{2} \dR     + 252 \zeta_{3}  \dR    - 36 \dR^2 \right)\\
       &+ \yt^6 \lambda   \left( - \f{1881}{8} \dR - 66 \zeta_{3} \dR   + 80 \dR^2 \right)
       + \yt^8   \left(\f{13}{2} \dR  - 12\zeta_{3} \dR   - \f{195}{8} \dR^2     \right)\\
       &+ \gs^2 \yt^2 \lambda^2   \left(  - 306 \cf \dR  + 288 \zeta_{3} \cf \dR   \right)
       + \gs^2 \yt^4 \lambda   \left(   \f{895}{4} \cf \dR  - 324 \zeta_{3} \cf \dR \right)\\
       &+ \gs^2 \yt^6   \left(  - \f{19}{2} \cf \dR    + 60 \zeta_{3} \cf \dR   \right)
       + \gs^4 \yt^2 \lambda   \left(  - \f{119}{2} \cf^2 \dR + 77 \ca \cf \dR   \right. \\ & \left.
                                       - 16 \Nf \tr \cf \dR   + 72 \zeta_{3} \cf^2 \dR    - 36 \zeta_{3} \ca \cf \dR  \right)
       + \gs^4 \yt^4   \left(   \f{131}{2} \cf^2 \dR     \right. \\ & \left.
     + 48 \tr \cf \dR           - \f{109}{2} \ca \cf \dR          + 10 \Nf \tr \cf \dR          - 48 \zeta_{3} \cf^2 \dR
          + 24 \zeta_{3} \ca \cf \dR     \right)
\end{split}
\label{beta:la}
{}.
\ee
The purely $\lambda$-dependent parts of eq.~(\ref{beta:la}) have been known for a while  
\cite{Brezin:1974xi,Brezin:1973}, the full one-loop
and two-loop result are in agreement with \cite{2loopbetayukawa,Machacek198570} (for an
SU$(3)$ colour gauge group). It has also been a useful check for our setup to
see that the same result can be derived from the four-$\Phi_1$ vertex,
the four-$\Phi_2$ vertex and the $(\Phi_1^{*} \Phi_2^{*}\Phi_1
\Phi_2)$ vertex.
\be
\begin{split}
\f{\beta_{\sss{\yt}}^{(1)}}{\yt}=&    \yt^2   \left(       \f{3}{4}     + \f{1}{2} \dR   \right)
       - 3 \cf\,\gs^2        ,\\    
\f{\beta_{\sss{\yt}}^{(2)}}{\yt}=&       
       3\, \lambda^2    
       -6\, \yt^2 \lambda      
       + \yt^4   \left(           \f{3}{4}          - \f{9}{4} \dR          \right)\\
       &+ \gs^2 \yt^2   \left(    6 \cf          + \f{5}{2} \cf \dR   \right)
       + \gs^4   \left( - \f{3}{2} \cf^2 - \f{97}{6} \ca \cf  + \f{10}{3} \Nf \tr \cf  \right),\\    
\f{\beta_{\sss{\yt}}^{(3)}}{\yt}=& 
       -18\, \lambda^3         
       + \yt^2 \lambda^2   \left(           \f{285}{8}          - \f{45}{4} \dR          \right)
       + \yt^4 \lambda   \left(           \f{63}{2}          + \f{45}{2} \dR          \right)\\
       &+ \yt^6   \left(          - \f{345}{32}          + \f{9}{4} \zeta_{3}          + \f{107}{32} \dR
          + \f{3}{2} \zeta_{3} \dR           + \f{39}{16} \dR^2          \right)\\
       &+ 6 \cf \, \gs^2 \yt^2 \lambda     
       - \gs^2 \yt^4   \left(    \f{57}{2} \cf      + \f{81}{8} \cf \dR    \right)\\
       &+ \gs^4 \yt^2   \left(
           \f{471}{16} \cf^2
          - \f{119}{8} \cf^2 \dR
          + 25 \tr \cf
          + \f{717}{16} \ca \cf
          + \f{77}{4} \ca \cf \dR \right. \\ &\left.
          - \f{33}{4} \Nf \tr \cf 
          - 4 \Nf \tr \cf \dR
          - 27 \zeta_{3} \cf^2
          + 18 \zeta_{3} \cf^2 \dR
          - \f{27}{2} \zeta_{3} \ca \cf \right. \\ &\left.
          - 9 \zeta_{3} \ca \cf \dR
          \right)
       + \gs^6   \left(
          - \f{129}{2} \cf^3
          + \f{129}{4} \ca \cf^2
          - \f{11413}{108} \ca^2 \cf
          + 46 \Nf \tr \cf^2             \right. \\ &\left.
          + \f{556}{27} \Nf \ca \tr \cf    
          + \f{140}{27} \Nf^2 \tr^2 \cf
          - 48 \zeta_{3} \Nf \tr \cf^2
          + 48 \zeta_{3} \Nf \ca \tr \cf
          \right)
{}.
\end{split}
\label{beta:yt}
\ee
The one-loop and two-loop part of this result have been found before in
\cite{2loopbetayukawa,Machacek1984221} (for $\dR=3$, $\tr=\f{1}{2}$) and the
contributions of order $\gs^2$, $\gs^4$, $\yt^2 \gs^2$ and $\gs^4 \yt$
to $Z_{\yt}$ have been successfully checked against
\cite{Steinhauser:1998cm}.  In this reference the calculation of
$Z_{M_t}$ has been performed in the broken phase of the SM with a massive top
quark. When comparing these two results one  has to take into account
that in the broken SM the top quark mass is to be
renormalized as a product $y_t\, (\Phi_2 + \Phi_2^{\dagger})$   so  the corresponding top 
quark mass renormalization constant is
\[
Z_{M_t} = Z_{\yt}\, Z_2^{(2\Phi)}=\f{Z^{(tb\Phi)}_1}{\sqrt{Z^{(2t)}_{2,L}Z^{(2t)}_{2,R}}} 
{} .
\]

Again the setup could be checked for consistency by using the renormalization
of the four different vertices t-t-$\Phi_2$, t-t-$\Phi_2^{*}$, 
t-b-$\Phi_1$, t-b-$\Phi_1^{*}$ for the calculation.
\be
\begin{split}
\f{\beta_{\sss{\gs}}^{(1)}}{\gs}=&   \gs^2   \left(  - \f{11}{6} \ca    + \f{2}{3} \Nf \tr  \right) ,\\
\f{\beta_{\sss{\gs}}^{(2)}}{\gs}=&     
       -  2 \tr \,\gs^2 \yt^2         
       + \gs^4   \left(    - \f{17}{3} \ca^2    + 2 \Nf \tr \cf + \f{10}{3} \Nf \ca \tr   \right),\\
\f{\beta_{\sss{\gs}}^{(3)}}{\gs}=&
        + \gs^2 \yt^4   \left(          \f{9}{2} \tr          + \f{7}{2}    \tr \dR          \right)
       - \gs^4 \yt^2   \left(           3 \tr \cf          + 12 \ca \tr          \right)\\
       &+ \gs^6   \left(
          - \f{2857}{108} \ca^3
          - \Nf \tr \cf^2
          + \f{205}{18} \Nf \ca \tr \cf
          + \f{1415}{54} \Nf \ca^2 \tr \right. \\ &\left.
          - \f{22}{9} \Nf^2 \tr^2 \cf
          - \f{79}{27} \Nf^2 \ca \tr^2
          \right)
{}.
\end{split}
\label{beta:gs}
\ee

In the case of $\yt=0$ this is in agreement with the well-known result \cite{Tarasov1980429,3loopbetaqcd}. The one-loop and two-loop parts of
eq.~(\ref{beta:gs}) are known from \cite{PhysRevLett.30.1343,PhysRevLett.30.1346,Fischler1982385,PhysRevD.25.581,Jack1985472,
Machacek198383},
the term $\propto \gs^4\yt^2$ can be found in \cite{Steinhauser:1998cm} and the full three-loop result has been computed
in \cite{PhysRevLett.108.151602} (for $\dR=3$, $\tr=\f{1}{2}$).\\ 
The $\beta$-function describing the running of the ``mass'' parameter $m^2$ in eq.~(\ref{LPhi}) can be computed from the renormalization constant of
the local operator $O_{2\Phi}:=\Phi^\dagger \Phi$. An insertion of $O_{2\Phi}$ into a Green's function, e.g. with two external $\Phi$-fields,
is renormalized as $[O_{2\Phi}]=Z_{\Phi^2}O_{2\Phi}$ where $[O_{2\Phi}]$ is the corresponding finite operator.
From $[O_{2\Phi}]=Z_{m^2}O^{\text{bare}}_{2\Phi}$ and $O^{\text{bare}}_{2\Phi}=Z_2^{(2\Phi)}O_{2\Phi}$ it follows that
\be Z_{m^2}=\lb Z_2^{(2\Phi)}\rb^{-1} Z_{\Phi^2}
{}. 
\ee 
This yields the following contributions to $\beta_{\sss{m^2}}$:
\be
\begin{split}
\f{\beta_{\sss{m^2}}^{(1)}}{m^2}=&   6\,     \lambda         
       +\dR\, \yt^2     ,\\
\f{\beta_{\sss{m^2}}^{(2)}}{m^2}=&   
         -30\, \lambda^2          
       -  12 \dR \,\yt^2 \lambda         
       - \f{9}{4} \dR\,\yt^4         
       +  5 \cf \dR\,\gs^2 \yt^2   ,\\
\f{\beta_{\sss{m^2}}^{(3)}}{m^2}=&   1026 \, \lambda^3 
       +   \f{99}{2} \dR \, \yt^2 \lambda^2 
       + \yt^4 \lambda   \left( \f{333}{4} \dR - 18 \dR^2 + 72 \zeta_{3} \dR \right)\\
       &+ \yt^6   \left(  - \f{617}{16} \dR + 24 \dR^2 + 15 \zeta_{3} \dR \right)\\
       &+ \gs^2 \yt^2 \lambda   \left(  - 153 \cf \dR + 144 \zeta_{3} \cf \dR \right)\\
       &+ \gs^2 \yt^4   \left( \f{447}{8} \cf \dR - 90 \zeta_{3} \cf \dR \right)\\
       &+ \gs^4 \yt^2   \left(  - \f{119}{4} \cf^2 \dR + \f{77}{2} \ca \cf \dR - 8 \Nf \tr \cf \dR + 36 
         \zeta_{3} \cf^2 \dR - 18 \zeta_{3} \ca \cf \dR \right)\\
{}.
\end{split}
\label{beta:m2}
\ee
The one-loop and two-loop parts of this result are in agreement with
\cite{2loopbetayukawa} where they have been computed before.
The purely $\lambda$-dependent part can be found in \cite{Brezin:1974xi,Brezin:1973}.
For $\dR=3$ and $\tr=\f{1}{2}$ (QCD) we get the following results:
\be
\begin{split}
\beta_{\sss{\lambda}}^{(1)}=&   12\, \lambda^2         + 6\,\yt^2 \lambda     - 3\,\yt^4          ,\\
\beta_{\sss{\lambda}}^{(2)}=&         - 156\,\lambda^3             
       - 72\, \yt^2 \lambda^2         
       - \f{3}{2}\, \yt^4 \lambda         
       + 15\,\yt^6            
       + 40\,\gs^2 \yt^2 \lambda           
       - 16\,\gs^2 \yt^4              ,\\   
\beta_{\sss{\lambda}}^{(3)}=& 
       \lambda^4   \left(           3588          + 2016 \zeta_{3}          \right)
       +  873 \,\yt^2 \lambda^3         
       + \yt^4 \lambda^2   \left(          \f{1719}{2}          + 756 \zeta_{3}          \right)\\
       &+ \yt^6 \lambda   \left(           \f{117}{8}          - 198 \zeta_{3}          \right)
       - \yt^8   \left(           \f{1599}{8}          + 36 \zeta_{3}          \right)
       + \gs^2 \yt^2 \lambda^2   \left(          - 1224          + 1152 \zeta_{3}          \right)\\
       &+ \gs^2 \yt^4 \lambda   \left(           895          - 1296 \zeta_{3}          \right)
       + \gs^2 \yt^6   \left(          - 38          + 240 \zeta_{3}          \right)\\
       &+ \gs^4 \yt^2 \lambda   \left(           \f{1820}{3}          - 32 \Nf          - 48 \zeta_{3}          \right)
       + \gs^4 \yt^4   \left(          - \f{626}{3}          + 20 \Nf
         + 32 \zeta_{3}          \right)
{}.
\end{split}
\ee
To get an idea of the size of these contributions and therefore the significance of our calculation we evaluate $\beta_{\sss{\lambda}}$ at the scale $\mu=M_Z$ 
(with an assumed Higgs mass of \mbox{125 GeV} and \mbox{$\Nf=6$}) which
yields a value of \mbox{$\beta_{\sss{\lambda}} \sim (-0.01)$} at one-loop level. 
The two and three-loop contributions change this result
by \mbox{$\sim 1\%$} and \mbox{$\sim (-0.04)\%$} respectively. To 
estimate the importance of the individual terms we introduce the labels
\be G=\f{\gs}{\gs(\mu=M_Z)},\quad Y=\f{\yt}{\yt(\mu=M_Z)},\quad L=\f{\lambda}{\lambda(\mu=M_Z)} 
\label{labels} 
\ee
and get
\be
\begin{split}
\beta_{\sss{\lambda}}|_{\mu=M_Z}=&
\lb \underbrace{-1.7 \,Y^4 }_{\text{\tiny 1 loop}}\rb 10^{-2}
+\lb \underbrace{5.0 \,L Y^2}_{\text{\tiny 1 loop}} 
\underbrace{+ 1.5\,  L^2 }_{\text{\tiny 1 loop}}\rb 10^{-3}\\
&+\lb \underbrace{-8.5\,  G^2 Y^4}_{\text{\tiny 2 loop}}
\underbrace{+5.0 \, Y^6}_{\text{\tiny 2 loop}} 
\underbrace{+ 3.1\,  G^2 L Y^2}_{\text{\tiny 2 loop}} \rb 10^{-4}\\
&+\lb \underbrace{7.9 \, G^2 Y^6}_{\text{\tiny 3 loop}} 
\underbrace{-4.8  \,Y^8}_{\text{\tiny 3 loop}} 
\underbrace{-5.3 \, L^2 Y^2}_{\text{\tiny 2 loop}} 
\underbrace{-3.1\,  G^2 L Y^4}_{\text{\tiny 3 loop}} \right.\\ &\left. 
\underbrace{-2.5\,  G^4 Y^4}_{\text{\tiny 3 loop}} 
\underbrace{+2.6 \, G^4 L Y^2}_{\text{\tiny 3 loop}} 
\underbrace{-1.7\,  L^3}_{\text{\tiny 2 loop}} \rb 10^{-5}\\
&+\lb \underbrace{-7.5\,  L Y^4}_{\text{\tiny 2 loop}}
\underbrace{+7.8 \, L^2 Y^4}_{\text{\tiny 3 loop}}
\underbrace{-6.6\,  L Y^6}_{\text{\tiny 3 loop}}
\underbrace{+1.1  \,G^2 L^2 Y^2}_{\text{\tiny 3 loop}}\rb 10^{-6}\\
&+\lb \underbrace{5.7\, L^3 Y^2}_{\text{\tiny 3 loop}}
\underbrace{+5.9 \, L^4}_{\text{\tiny 3 loop}} \rb 10^{-7}
{}.
\end{split}
\ee
We see that the decrease of the effective four-$\Phi$ coupling with
increasing energy is induced by top quark loops. Without quarks there would be an increase.
It is also worth noting that the individual contributions at three-loop level are much larger than the overall effect 
due to huge cancellations. Consider for example the five numerically largest three-loop terms at $\mu=M_Z$:
$$\lb 7.9 \, G^2 Y^6 -4.8  \,Y^8 -3.1\,  G^2 L Y^4 -2.5\,  G^4 Y^4 +2.6 \, G^4 L Y^2 \rb 10^{-5}.$$
The total contribution from these terms is by almost two orders of magnitude smaller than the size of the largest 
one.\\
For the top-Yukawa $\beta$-functions we find
\be
\begin{split}
\f{\beta_{\sss{\yt}}^{(1)}}{\yt}=&  \f{9}{4}\,       \yt^2              -4\, \gs^2        ,\\
\f{\beta_{\sss{\yt}}^{(2)}}{\yt}=&        3\,     \lambda^2            
       -6\, \yt^2 \lambda       -6\, \yt^4        +18\, \gs^2 \yt^2      
       + \gs^4   \left(          - \f{202}{3}          + \f{20}{9} \Nf          \right) ,\\   
\f{\beta_{\sss{\yt}}^{(3)}}{\yt}=& 
       - 18\,\lambda^3     
       +  \f{15}{8} \,\yt^2 \lambda^2         
       +99\, \yt^4 \lambda     
       + \yt^6   \left(     \f{339}{16}          + \f{27}{4} \zeta_{3}          \right)\\
       &+8\, \gs^2 \yt^2 \lambda          
       - \f{157}{2}\,\gs^2 \yt^4     
       + \gs^4 \yt^2   \left(    \f{4799}{12}          - \f{27}{2} \Nf          - 114 \zeta_{3}    \right)\\
       &+ \gs^6   \left(  - 1249   + \f{2216}{27} \Nf   + \f{140}{81}
         \Nf^2   + \f{160}{3} \zeta_{3} \Nf \right)
{}.
\end{split}
\ee
As has already been mentioned above the $\lambda$-corrections are negligible here. Evaluating
$\beta_{\sss{\yt}}$ at the scale $\mu=M_Z$ (with an assumed Higgs mass of 125 GeV and $\Nf=6$) we get a value of $\sim(-0.023)$ at 
one-loop level which means a decrease of $\yt$ and therefore the top mass with increasing energy. This is due to the QCD corrections.
In the absence of QCD the opposite would be the case as we can see from
the term $\propto \yt^2$ in $\f{\beta_{\sss{\yt}}}{\yt}$. The
two and three-loop corrections are $\sim 16.6\%$ and $\sim 0.7\%$ with respect to the one-loop result and
so quite high compared e.g. to the case of $\beta_{\sss{\gs}}$ discussed below. We use again the labels \eqref{labels}
to get an impression of the individual terms:
\be
\begin{split}
\beta_{\sss{\yt}}|_{\mu=M_Z}=&
\lb \underbrace{-3.7 \, Y G^2}_{\text{\tiny 1 loop}} 
\underbrace{+1.3 \, Y^3}_{\text{\tiny 1 loop}} \rb 10^{-2}
+\lb \underbrace{-4.7 \, Y G^4}_{\text{\tiny 2 loop}}  \rb 10^{-3}\\
&+\lb \underbrace{9.8 \, Y^3 G^2}_{\text{\tiny 2 loop}}
\underbrace{-2.5 \, Y G^6}_{\text{\tiny 3 loop}} 
\underbrace{-2.1 \, Y^5}_{\text{\tiny 2 loop}} \rb 10^{-4}\\
&+\lb \underbrace{9.3 \, Y^3 G^4}_{\text{\tiny 3 loop}}
\underbrace{-3.1 \, L Y^3}_{\text{\tiny 2 loop}}
\underbrace{-2.5 \, Y^5 G^2}_{\text{\tiny 3 loop}} \rb 10^{-5}\\
&+\lb \underbrace{6.0 \, Y^7}_{\text{\tiny 3 loop}} 
\underbrace{+3.0 \, L Y^5}_{\text{\tiny 3 loop}} 
\underbrace{+2.3 \, L^2 Y}_{\text{\tiny 2 loop}}\rb 10^{-6}\\
&+\lb \underbrace{3.9 \, L Y^3 G^2}_{\text{\tiny 3 loop}} \rb 10^{-7}
+\lb \underbrace{-1.2 \, L^3 Y}_{\text{\tiny 3 loop}} \rb 10^{-8}
+\lb \underbrace{8.5 \, L^2 Y^3}_{\text{\tiny 3 loop}} \rb 10^{-9}
{}.
\end{split}
\ee
For the strong coupling we get

\be
\begin{split}
\f{\beta_{\sss{\gs}}^{(1)}}{\gs}=&  \gs^2   \left(          - \f{11}{2}          + \f{1}{3} \Nf          \right)  ,\\   
\f{\beta_{\sss{\gs}}^{(2)}}{\gs}=&          - \gs^2 \yt^2   
       + \gs^4   \left(          - 51          + \f{19}{3} \Nf          \right) ,\\  
\f{\beta_{\sss{\gs}}^{(3)}}{\gs}=&     \f{15}{2}\,     \gs^2 \yt^4           -20\, \gs^4 \yt^2     
       + \gs^6   \left(  - \f{2857}{4}   + \f{5033}{36} \Nf     -
         \f{325}{108} \Nf^2   \right)
{}.
\end{split}
\ee
In order to numerically compare the higher order corrections to the above $\beta$-functions we also give the evaluation of
$\beta_{\sss{\gs}}$ at the scale $\mu=M_Z$ and with $\Nf=6$. The one-loop contribution is $\sim (-0.04)$ to which the two and three-loop calculations
give corrections of $\sim 3.7\%$ and $\sim (-0.02)\%$ respectively. Very small $\lambda$-corrections to $\beta_{\sss{\gs}}$ do not appear
until four loops. With eq.~(\ref{labels}) and the above assumptions we get
\be
\begin{split}
\beta_{\sss{\gs}}|_{\mu=M_Z}=&
\lb\underbrace{-4.0\, G^3}_{\text{\tiny 1 loop}}\rb 10^{-2}
+\lb\underbrace{-1.4\, G^5}_{\text{\tiny 2 loop}}\rb 10^{-3}\\
&+\lb\underbrace{-6.9\, Y^2 G^3}_{\text{\tiny 2 loop}} \underbrace{+ 1.7\, G^7}_{\text{\tiny 3 loop}}
 \underbrace{-1.3 \,Y^2 G^5 }_{\text{\tiny 3 loop}}\rb 10^{-5}
+\lb \underbrace{3.1 \,Y^4 G^3}_{\text{\tiny 3 loop}} \rb 10^{-6}
{}.
\end{split}
\ee

The running of the $m^2$ parameter is given by
\be
\begin{split}
\f{\beta_{\sss{m^2}}^{(1)}}{m^2}=& 
       6\, \lambda          
       +3\, \yt^2        ,\\
\f{\beta_{\sss{m^2}}^{(2)}}{m^2}=&  
          -30\, \lambda^2             
       - 36\, \yt^2 \lambda         
       - \f{27}{4} \, \yt^4      
       + 20\,\gs^2 \yt^2       ,\\ 
\f{\beta_{\sss{m^2}}^{(3)}}{m^2}=&       1026 \,\lambda^3  
       + \f{297}{2}\, \yt^2 \lambda^2   
       + \yt^4 \lambda   \left(
          \f{351}{4}
          + 216 \zeta_{3}
          \right)\\
       &+ \yt^6   \left(
           \f{1605}{16}
          + 45 \zeta_{3}
          \right)
       + \gs^2 \yt^2 \lambda   \left(
          - 612
          + 576 \zeta_{3}
          \right)\\
       &+ \gs^2 \yt^4   \left(
           \f{447}{2}
          - 360 \zeta_{3}
          \right)
       + \gs^4 \yt^2   \left(
           \f{910}{3}
          - 16 \Nf
          - 24 \zeta_{3}
          \right)      
{}.
\end{split}
\ee
Again we evaluate $\beta_{\sss{m^2}}$ at the scale $\mu=M_Z$ (with an assumed Higgs mass of 125 GeV and $\Nf=6$) 
and get a value of $\sim(+0.023)$ at one-loop level which means an increase of $m^2$ at higher energy scales. 
The two and three-loop corrections are $\sim 2.9\%$ and $\sim 0.32\%$. With the labels \eqref{labels} we can estimate
the contributions of the individual terms:
\be
\begin{split}
\f{\beta_{\sss{m^2}}}{m^2}|_{\mu=M_Z}=&
\lb \underbrace{1.8\, Y^2}_{\text{\tiny 1 loop}}\rb 10^{-2}
+\lb \underbrace{5.3\, L}_{\text{\tiny 1 loop}}
\underbrace{+1.1 \,G^2 Y^2}_{\text{\tiny 2 loop}}\rb 10^{-3}
+\lb \underbrace{-2.4\, Y^4}_{\text{\tiny 2 loop}}
\underbrace{-1.9\, L Y^2}_{\text{\tiny 2 loop}}\rb 10^{-4}\\
&+\lb
\underbrace{9.4\, G^4 Y^2}_{\text{\tiny 3 loop}}
\underbrace{-7.0\, G^2 Y^4}_{\text{\tiny 3 loop}}
\underbrace{+3.3\, Y^6}_{\text{\tiny 3 loop}}
\underbrace{-2.4 \, L^2}_{\text{\tiny 2 loop}}
\underbrace{+1.1\, L Y^4}_{\text{\tiny 3 loop}}
\rb 10^{-5}\\
&+\lb 
\underbrace{4.0\, G^2 L Y^2}_{\text{\tiny 3 loop}}
\rb 10^{-6}
+\lb \underbrace{7.1\, L^3}_{\text{\tiny 3 loop}}
\underbrace{+7.0\, L^2 Y^2}_{\text{\tiny 3 loop}}\rb 10^{-7}
{}.
\end{split}
\ee

\section{Results for the anomalous dimensions \label{res:AD}}
In this section we give the anomalous dimensions of the physical fields in this
setup. Note that because of the SU$(2)$ symmetry $\Phi_1$ and $\Phi_2$
must have the same anomalous dimension $\gamma_{2}^{\sss{\Phi}}$. The same holds for the
left-handed part of top and bottom quarks: \mbox{$\gamma_{2,L}^{\sss{t}}=\gamma_{2,L}^{\sss{b}}$.}
For the quark flavours q which do not participate in the Yukawa
interaction there is no difference between the left- and right-handed
part as they are renormalized by the same Z-factor:
\mbox{$\gamma_{2}^{\sss{q}}\equiv\gamma_{2,L}^{\sss{q}}=\gamma_{2,R}^{\sss{q}}$.}
This also applies to the right-handed part of the bottom quark:
\mbox{$\gamma_{2,R}^{\sss{b}}=\gamma_{2}^{\sss{q}}$.} All these relations have been
tested explicitly during our calculation which provides a nice additional
check.

\be
\begin{split}
\gamma_2^{{\sss{q}}\,(1)}=&  \gs^2  \, \cf \left(           1         - \xi          \right),\\
\gamma_2^{{\sss{q}}\,(2)}=&  \gs^4   \left(
          - \f{3}{2} \cf^2
          + \f{17}{2} \ca \cf
          - 2 \Nf \tr \cf
          - \f{5}{2} \xi \ca \cf
          + \f{1}{4} \xi^2 \ca \cf
          \right),\\
\gamma_2^{{\sss{q}}\,(3)}=& 6 \tr \cf \, \gs^4 \yt^2   
       + \gs^6   \left(
           \f{3}{2} \cf^3
          - \f{143}{4} \ca \cf^2
          + \f{10559}{144} \ca^2 \cf       
          + 3 \Nf \tr \cf^2                 \right. \\ &\left. 
          - \f{1301}{36} \Nf \ca \tr \cf
          + \f{20}{9} \Nf^2 \tr^2 \cf
          + 12 \zeta_{3} \ca \cf^2
          - \f{15}{2} \zeta_{3} \ca^2 \cf   
          - \f{371}{32} \xi \ca^2 \cf        \right. \\ &\left.
          + \f{17}{4} \xi \Nf \ca \tr \cf
          - \f{3}{2} \xi \zeta_{3} \ca^2 \cf
          + \f{69}{32} \xi^2 \ca^2 \cf
          + \f{3}{8} \xi^2 \zeta_{3} \ca^2 \cf
          - \f{5}{16} \xi^3 \ca^2 \cf
          \right)        
{}.
\end{split}
\ee
For $\yt=0$ this is in agreement with the well-known QCD result \cite{3loopbetaqcd}.
The renormalization constants for t,b and q can also be found in \cite{Steinhauser:1998cm}
up to order $\gs^6$ and $\gs^4\yt^2$ where the calculation has been performed in the broken phase of the SM.
\be
\begin{split}
\gamma_{2,L}^{{\sss{t}}\,(1)}=& \gamma_2^{{\sss{q}}\,(1)} 
       + \f{1}{2} \, \yt^2  ,              \\
\gamma_{2,L}^{{\sss{t}}\,(2)}=& \gamma_2^{{\sss{q}}\,(2)} 
     - \yt^4   \left(           \f{1}{4}          + \f{3}{4} \dR          \right)
     - 2 \cf\, \gs^2 \yt^2  , \\
\gamma_{2,L}^{{\sss{t}}\,(3)}=& \gamma_2^{{\sss{q}}\,(3)}   
    - \f{33}{4}\, \yt^2 \lambda^2  
       +6\, \yt^4 \lambda   
       + \yt^6   \left(
          - \f{3}{2}
          + \f{29}{8} \dR
          - \f{3}{8} \dR^2
          + \f{3}{2} \zeta_{3}
          \right)\\ &
       + \gs^2 \yt^4   \left(
           \f{13}{2} \cf
          + \f{5}{8} \cf \dR
          + 6 \zeta_{3} \cf \dR
          \right)\\ &
       + \gs^4 \yt^2   \left(
          - \f{51}{8} \cf^2
          + \f{31}{8} \ca \cf
          - \f{3}{2} \Nf \tr \cf
          + 6 \zeta_{3} \cf^2
          - 15 \zeta_{3} \ca \cf
          \right)
{}.
\end{split}
\ee
\be
\begin{split}
\gamma_{2,R}^{{\sss{t}}\,(1)}=& \gamma_2^{{\sss{q}}\,(1)}     + \yt^2     ,   \\
\gamma_{2,R}^{{\sss{t}}\,(2)}=& \gamma_2^{{\sss{q}}\,(2)}    
 - \yt^4   \left(           \f{1}{4}          + \f{3}{2} \dR          \right)        - 4 \cf \, \gs^2 \yt^2    ,   \\
\gamma_{2,R}^{{\sss{t}}\,(3)}=& \gamma_2^{{\sss{q}}\,(3)}  
  - \f{33}{2} \, \yt^2 \lambda^2   
       + 12\, \yt^4 \lambda  
       + \yt^6   \left(          - \f{33}{16}      + \f{53}{8} \dR   - \f{3}{4} \dR^2   + 3 \zeta_{3}    \right) \\ &
       + \gs^2 \yt^4   \left(
           5 \cf
          + \f{5}{4} \cf \dR
          + 12 \zeta_{3} \cf \dR
          \right) \\ &
       + \gs^4 \yt^2   \left(
          - \f{51}{4} \cf^2
          + \f{31}{4} \ca \cf
          - 3 \Nf \tr \cf
          + 12 \zeta_{3} \cf^2
          - 30 \zeta_{3} \ca \cf
          \right)
{}.
\end{split}
\ee
\be
\begin{split}
\gamma_2^{{\sss{\Phi}}\,(1)}=&  \dR \, \yt^2   , \\
\gamma_2^{{\sss{\Phi}}\,(2)}=&     6\, \lambda^2         - \f{9}{4} \dR \, \yt^4         +  5 \cf \dR \, \gs^2 \yt^2 , \\
\gamma_2^{{\sss{\Phi}}\,(3)}=&     - 36\,  \lambda^3   
       - \f{45}{2} \dR \, \yt^2 \lambda^2   
       + 15 \dR \, \yt^4 \lambda  \\ &
       + \yt^6   \left(          - \f{25}{16} \dR          + 6 \dR^2          + 3 \zeta_{3} \dR          \right)
       + \gs^2 \yt^4   \left(          \f{15}{8} \cf \dR          - 18 \zeta_{3} \cf \dR          \right)\\ &
       + \gs^4 \yt^2   \left(
          - \f{119}{4} \cf^2 \dR
          + \f{77}{2} \ca \cf \dR
          - 8 \Nf \tr \cf \dR
          + 36 \zeta_{3} \cf^2 \dR
          - 18 \zeta_{3} \ca \cf \dR
          \right)
{},
\end{split}
\ee
The purely $\lambda$-dependent part of this has been computed before in \cite{Brezin:1974xi,Brezin:1973}.
\be
\begin{split}
\gamma_2^{{\sss{g}}\,(1)}=& \gs^2   \left(     - \f{5}{3} \ca     + \f{4}{3} \Nf \tr      - \f{1}{2} \xi \ca     \right)   ,  \\
\gamma_2^{{\sss{g}}\,(2)}=& -4\tr \, \gs^2 \yt^2  
       + \gs^4   \left(
          - \f{23}{4} \ca^2
          + 4 \Nf \tr \cf
          + 5 \Nf \ca \tr
          - \f{15}{8} \xi \ca^2
          + \f{1}{4} \xi^2 \ca^2
          \right)   , \\
\gamma_2^{{\sss{g}}\,(3)}=&    \gs^2 \yt^4   \left(           9 \tr          + 7 \tr \dR          \right)
       - \gs^4 \yt^2   \left(           6 \tr \cf          + \f{25}{2} \ca \tr          \right) \\ &
       + \gs^6   \left(
          - \f{4051}{144} \ca^3
          - 2 \Nf \tr \cf^2
          + \f{5}{18} \Nf \ca \tr \cf
          + \f{875}{18} \Nf \ca^2 \tr
          - \f{44}{9} \Nf^2 \tr^2 \cf \right. \\ & \left.
          - \f{76}{9} \Nf^2 \ca \tr^2  
          + \f{3}{2} \zeta_{3} \ca^3
          + 24 \zeta_{3} \Nf \ca \tr \cf
          - 18 \zeta_{3} \Nf \ca^2 \tr
          - \f{127}{16} \xi \ca^3 \right. \\ & \left.
          + 2 \xi \Nf \ca^2 \tr
          - \f{9}{8} \xi \zeta_{3} \ca^3
          + \f{27}{16} \xi^2 \ca^3
          + \f{3}{16} \xi^2 \zeta_{3} \ca^3
          - \f{7}{32} \xi^3 \ca^3
          \right) 
{}.
\\
\end{split}
\ee
This is also in agreement with \cite{3loopbetaqcd} for $\yt=0$ and with \cite{Tarasov1980429} for $\yt=0$, $\xi=0$.
For $\dR=3$ and $\tr=\f{1}{2}$ (QCD) these results are as follows:
\be
\begin{split}
\gamma_2^{{\sss{q}}\,(1)}=&         \f{4}{3}  \left(         1- \xi          \right) \,\gs^2 ,\\
\gamma_2^{{\sss{q}}\,(2)}=&     \gs^4   \left(
           \f{94}{3}
          - \f{4}{3} \Nf
          - 10 \xi
          + \xi^2
          \right),\\
\gamma_2^{{\sss{q}}\,(3)}=& 4\, \gs^4 \yt^2  
       + \gs^6   \left(
          + \f{24941}{36}
          - \f{1253}{18} \Nf
          + \f{20}{27} \Nf^2
          - 26 \zeta_{3}
          - \f{1113}{8} \xi \right. \\ & \left.
          + \f{17}{2} \xi \Nf
          - 18 \xi \zeta_{3}
          + \f{207}{8} \xi^2
          + \f{9}{2} \xi^2 \zeta_{3}
          - \f{15}{4} \xi^3
          \right)      
{},
\end{split}
\ee

\be
\begin{split}
\gamma_{2,L}^{{\sss{t}}\,(1)}=& \gamma_2^{{\sss{q}}\,(1)} 
       + \f{1}{2} \, \yt^2      ,  \\
\gamma_{2,L}^{{\sss{t}}\,(2)}=& \gamma_2^{{\sss{q}}\,(2)} 
        - \f{5}{2} \,  \yt^4   
        - \f{8}{3} \,  \gs^2 \yt^2   ,   \\
\gamma_{2,L}^{{\sss{t}}\,(3)}=& \gamma_2^{{\sss{q}}\,(3)}  
- \f{33}{4}   \, \yt^2 \lambda^2  
       + 6\, \yt^4 \lambda   
       + \yt^6   \left(           6          + \f{3}{2} \zeta_{3}          \right) \\ &
       + \gs^2 \yt^4   \left(           \f{67}{6}          + 24 \zeta_{3}          \right)
       + \gs^4 \yt^2   \left(           \f{25}{6}          - \Nf          - \f{148}{3} \zeta_{3}          \right)    
{}.
\end{split}
\ee

\be
\begin{split}
\gamma_{2,R}^{{\sss{t}}\,(1)}=& \gamma_2^{{\sss{q}}\,(1)}     
       + \yt^2     ,  \\
\gamma_{2,R}^{{\sss{t}}\,(2)}=& \gamma_2^{{\sss{q}}\,(2)}    
       - \f{19}{4} \, \yt^4   
       - \f{16}{3} \, \gs^2 \yt^2     ,\\
\gamma_{2,R}^{{\sss{t}}\,(3)}=& \gamma_2^{{\sss{q}}\,(3)}  
       - \f{33}{2} \yt^2 \lambda^2  
       + 12\, \yt^4 \lambda   
       + \yt^6   \left(          \f{177}{16}          + 3 \zeta_{3}          \right)\\ &
       + \gs^2 \yt^4   \left(           \f{35}{3}          + 48 \zeta_{3}          \right)
       + \gs^4 \yt^2   \left(           \f{25}{3}          - 2 \Nf          - \f{296}{3} \zeta_{3}          \right)       
{},
\end{split}
\ee

\be
\begin{split}
\gamma_2^{{\sss{\Phi}}\,(1)}=&  3\,    \yt^2   ,  \\
\gamma_2^{{\sss{\Phi}}\,(2)}=&  6\,    \lambda^2  
       - \f{27}{4}\,  \yt^4   
       + 20\, \gs^2 \yt^2  ,  \\
\gamma_2^{{\sss{\Phi}}\,(3)}=&  - 36 \,     \lambda^3  
       - \f{135}{2} \, \yt^2 \lambda^2   
       + 45 \, \yt^4 \lambda   
       + \yt^6   \left(           \f{789}{16}          + 9 \zeta_{3}          \right)\\ &
       + \gs^2 \yt^4   \left(           \f{15}{2}          - 72 \zeta_{3}          \right)
       + \gs^4 \yt^2   \left(           \f{910}{3}          - 16 \Nf          - 24 \zeta_{3}          \right)     
{},
\end{split}
\ee

\be
\begin{split}
\gamma_2^{{\sss{g}}\,(1)}=&    \gs^2   \left(          - 5          + \f{2}{3} \Nf          - \f{3}{2} \xi          \right)  ,   \\
\gamma_2^{{\sss{g}}\,(2)}=&    -2 \, \gs^2 \yt^2  
       + \gs^4   \left(
          - \f{207}{4}
          + \f{61}{6} \Nf
          - \f{135}{8} \xi
          + \f{9}{4} \xi^2
          \right)  ,  \\
\gamma_2^{{\sss{g}}\,(3)}=& 15\,    \gs^2 \yt^4  
       - \f{91}{4}  \, \gs^4 \yt^2  \\ &
       + \gs^6   \left(
          - \f{12153}{16}
          + \f{7831}{36} \Nf
          - \f{215}{27} \Nf^2
          + \f{81}{2} \zeta_{3}
          - 33 \zeta_{3} \Nf
          - \f{3429}{16} \xi \right. \\ & \left.
          + 9 \xi \Nf
          - \f{243}{8} \xi \zeta_{3}
          + \f{729}{16} \xi^2
          + \f{81}{16} \xi^2 \zeta_{3}
          - \f{189}{32} \xi^3
          \right)     
{}.
\end{split}
\ee

\section{The evolution of the quartic Higgs coupling \label{la:evolution}}

The   quartic Higgs coupling $\lambda$  is of
special interest as it is directly  related to the Higgs mass $M_H$. 
If we assume that the SM is valid up to some  high energy scale  $\Lambda$, then
the value of $M_H$ should meet the constraints 
\[
m_{min}  < M_{\sss{H}} < m_{max}   
{}.
\]
Here the upper limit is related to the well-known fact that the running
Higgs self-coupling develops a Landau pole\footnote{This is true in the
one-loop approximation. At two loops  the Landau pole is replaced by 
an ultraviolet  metastable fixed point with the resulting fixed point value of
$\lambda$ being outside the weak coupling region.} if $M_H$ is large \cite{Maiani:1977cg,Cabibbo:1979ay,Lindner:1985uk}.
For $\Lambda = M_{Planck} = 10^{18}$ GeV the estimated value of $ m_{max} $ is around 175 GeV 
\cite{Maiani:1977cg,Cabibbo:1979ay,Lindner:1985uk,Hambye:1996wb}, which is already  
excluded by experiments carried out at the LHC and the Tevatron.

The lower limit $m_{min}$ follows from the requirement of the vacuum
stability  \cite{Krasnikov:1978pu,Hung:1979dn,Politzer:1978ic}. In 
order to find $m_{min}$  one should
construct the effective Higgs potential $V[\phi]$  including radiative
corrections and sum possible large logarithms using the standard
method of the Renormalization Group (for a review see, e.g. \cite{Sher:1988mj}). 
Once this has been done, the condition that the potential  $V[\phi]$ 
does not develop a deeper minimum in addition to the standard one  
for all values of $\phi < \Lambda $  fixes $m_{min}$.

In our analysis we will use a simplified approach for finding
$m_{min}$, namely the requirement that the running coupling constant
$\lambda(\mu)$ stay non-negative for all $\mu$ less than
$\Lambda$. It has been shown in \cite{Cabibbo:1979ay,Ford:1992mv} that the
simplified approach is essentially equivalent to the one based on the
use of the effective potential provided the instabilty of $V[\phi]$ can
only happen at $\phi \gg M_Z $.

In this section  we investigate the
effect of the three-loop result $\beta_{\sss{\lambda}}^{(3)}$ on the
running of $\lambda$ and therefore its effect on the stability of the
electroweak vacuum in the SM.  For this we also include the
electroweak contributions up to the two-loop level. The two-loop
$\beta$-functions for the SM gauge couplings have been derived in
\cite{Fischler1982385,PhysRevD.25.581,Jack1985472,
Machacek198383}. The two-loop results for the Yukawa-couplings and
$\lambda$ can be found in \cite{2loopbetayukawa, Machacek1984221,
Machacek198570}. Now we add the three-loop results derived in the
previous section and investigate the effect this has on the evolution
of our couplings (for a recent similar analysis, using the two-loop running, see, e.g. 
\cite{EliasMiro:2011aa,Xing:2011aa,Holthausen:2011aa}).

To find starting values for the running of the
couplings we should  account for the fact that the physical parameters
(e.g. pole masses) are related to the ones in the
$\overline{\text{MS}}$-scheme in a non-trivial way (see
e.g. \cite{Espinosa:2007qp,Hempfling:1994ar, Sirlin1986389}). For
example eq.~(\ref{lambdatree}) is only valid at tree level. For the
higher order corrections we take the electroweak ones at one-loop
and the QCD ones at two-loop level from
\cite{Espinosa:2007qp,Hempfling:1994ar}.  These matching relations  depend on
the exact values of $\als(M_Z)$ and the pole mass $M_{{t}}$ of
the top quark and of course the mass of the Higgs boson
$M_H$. For the latter we consider the cases
\mbox{$M_H=124$ GeV} and \mbox{$M_H=126$ GeV.}  For the other
two we use the values
\be
 \als(M_Z)=0.1184 \pm 0.0007 \qquad \text{and} \qquad M_{{t}}
=172.9 \pm 0.6 \pm 0.9 \text{ GeV} \quad \text{\cite{pdg}}
{}. 
\label{alsmt}
\ee
One should also keep in mind that the matching relations themselves receive 
contributions from  not yet known higher order corrections . The corresponding 
uncertainty in $m_{min}$  has been estimated in \cite{Bezrukov:2009db,EliasMiro:2011aa}
and found to be about $2$ GeV.

\begin{figure}[h!]
\includegraphics[width=\textwidth]{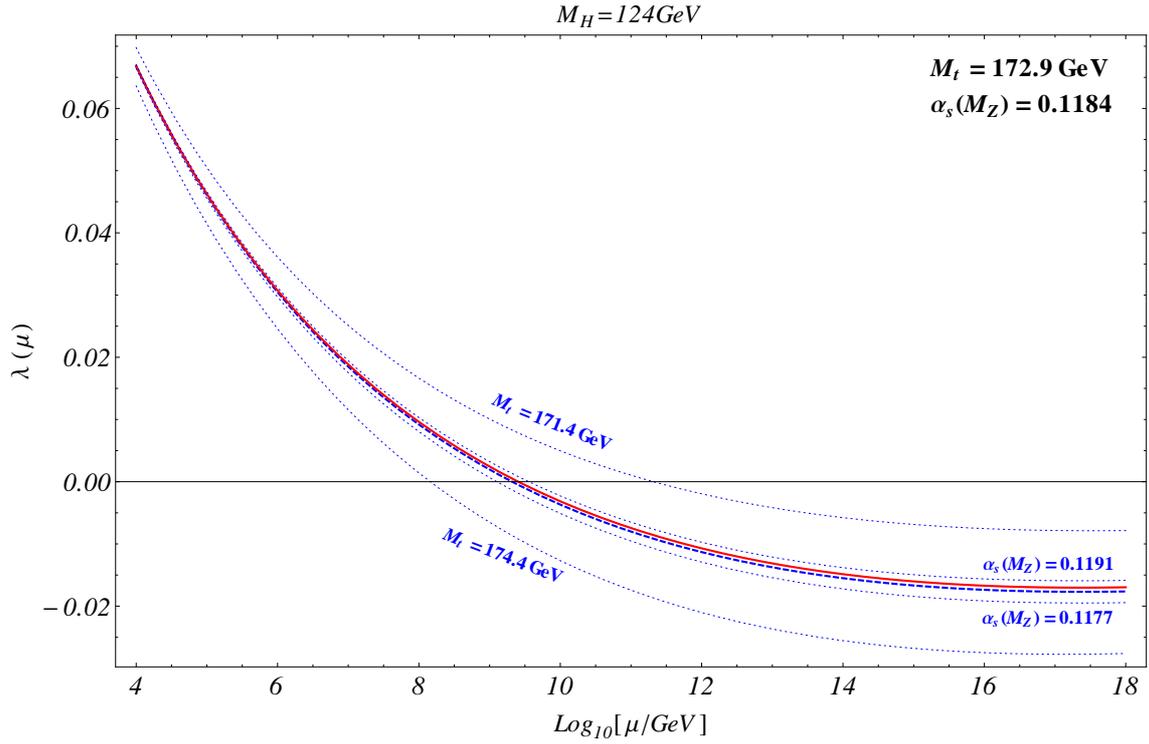} 
\vskip 0.5cm
\includegraphics[width=\textwidth]{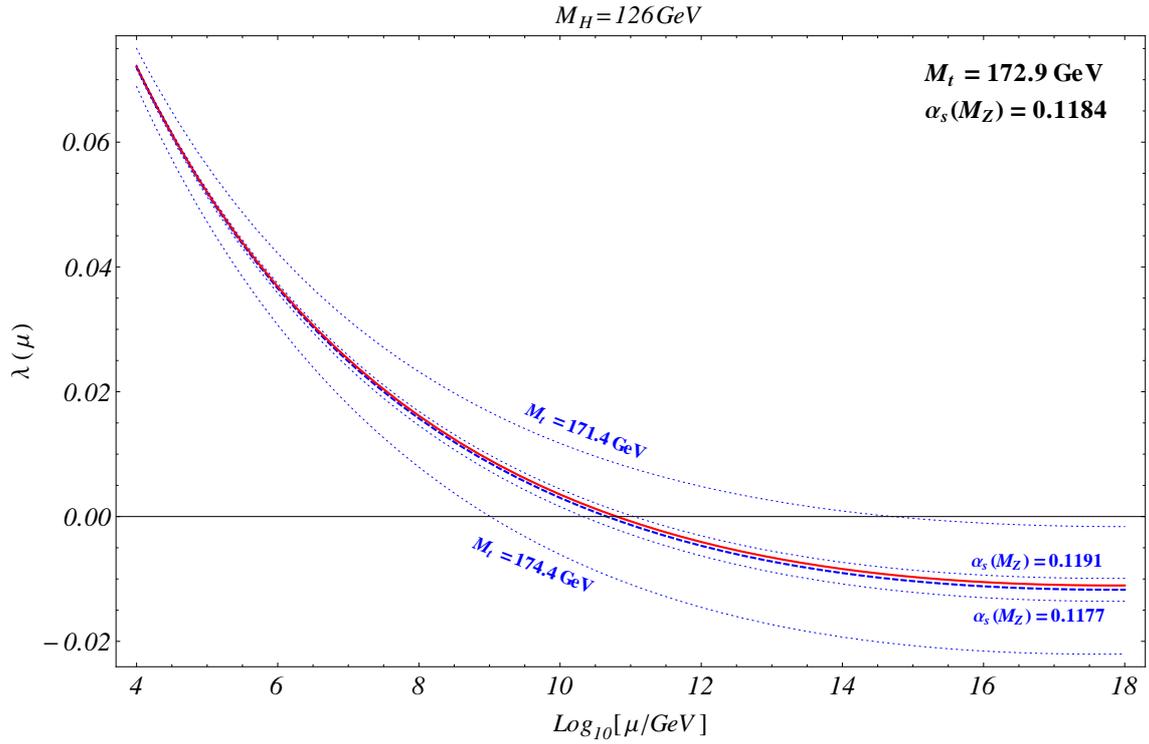} 
\caption{Evolution of $\lambda$ with the scale $\mu$: \Blue{2 loop (dashed, blue)} and \Red{3 loop (continuous, red)} results;
 Uncertainties with respect to the two-loop result: $\pm 1\sigma_{\als}$, $\pm 1\sigma_{M_t}$ (dotted)}
\label{lambda124126}
\end{figure}

\begin{figure}[h!]
\includegraphics[width=\textwidth]{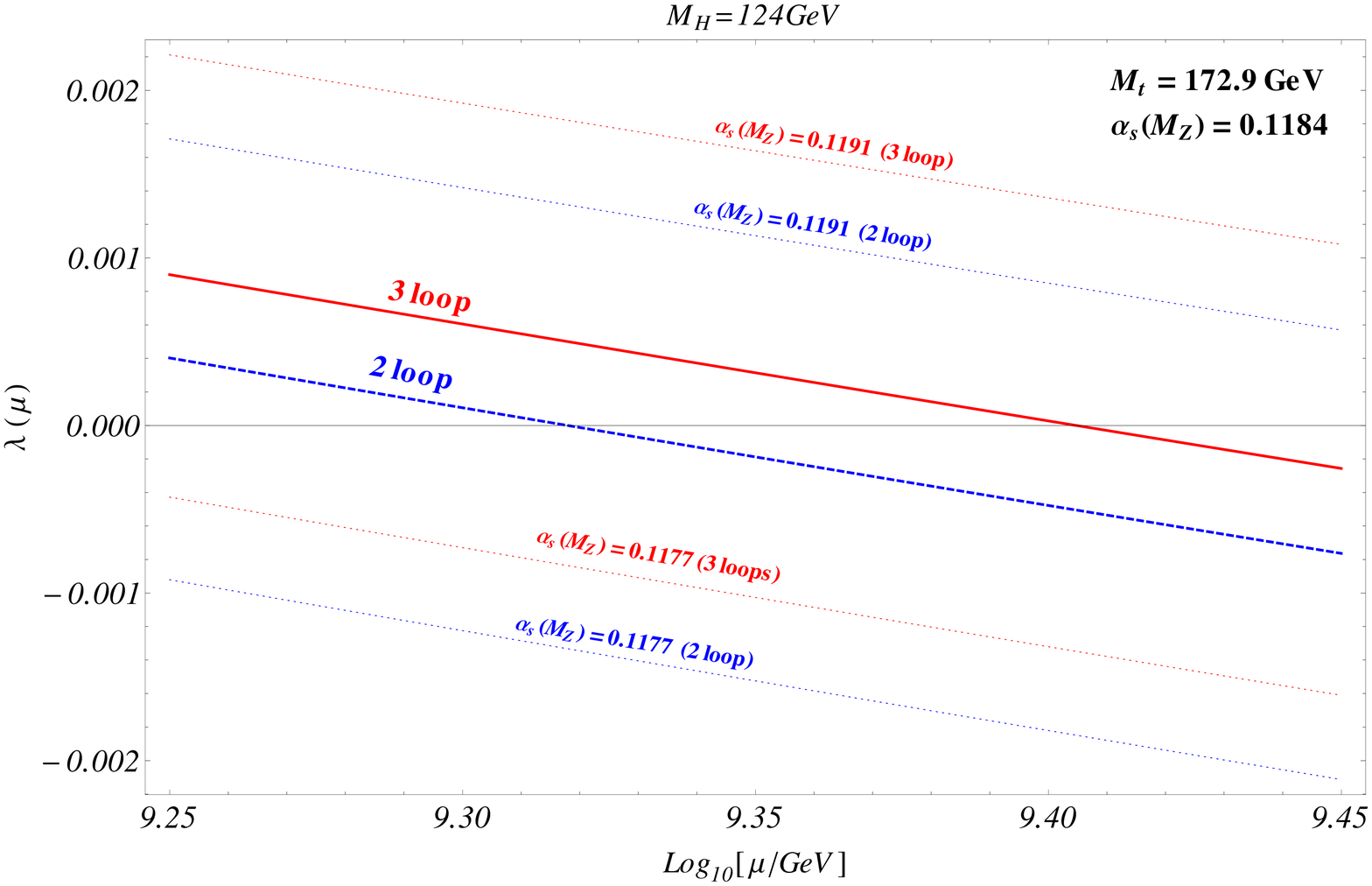} 
\vskip 1cm
\includegraphics[width=\textwidth]{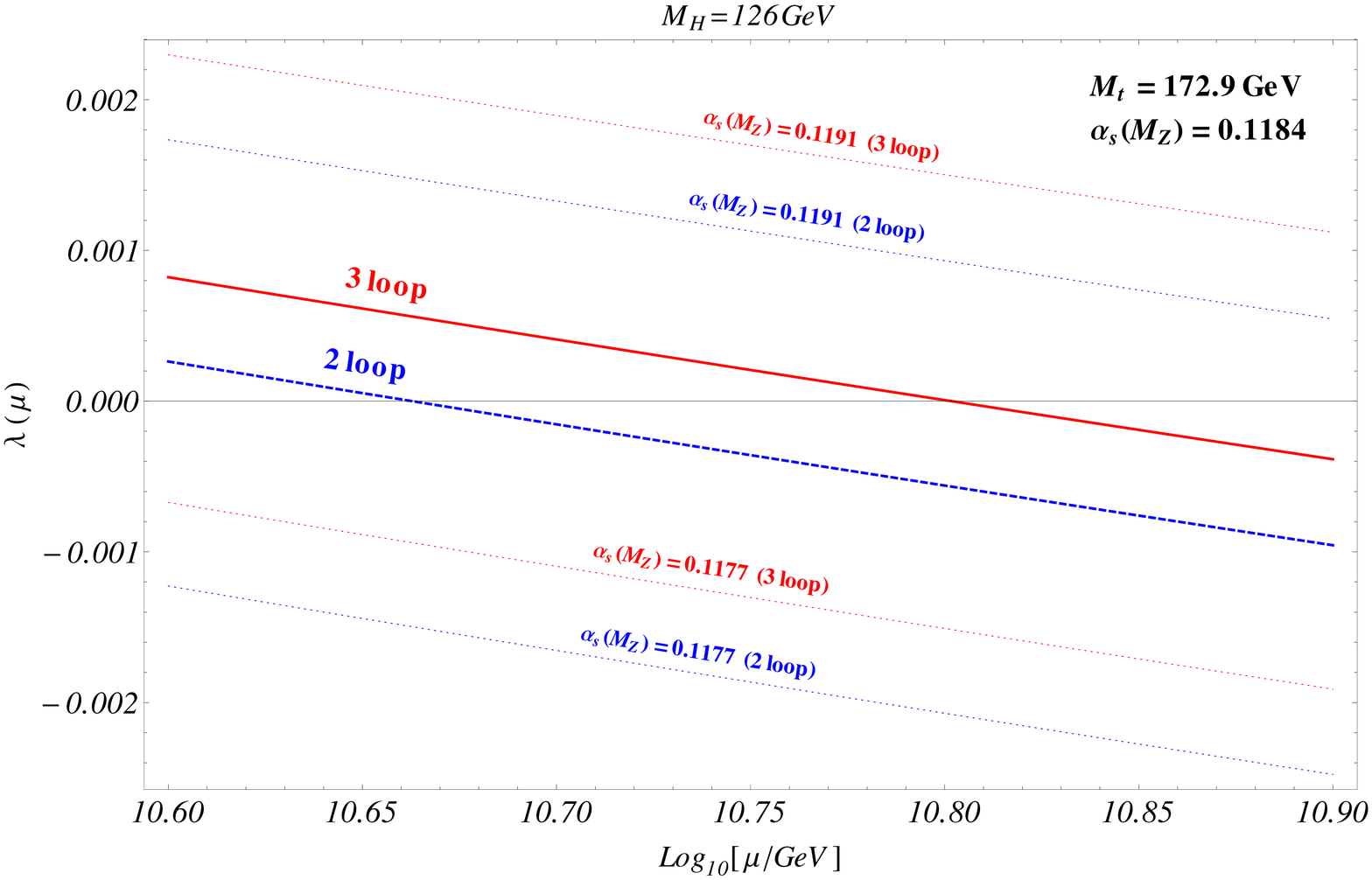} 
\caption{Evolution of $\lambda$ with the scale $\mu$: \Blue{2 loop (dashed, blue)} and \Red{3 loop (continuous, red)} results;
 Uncertainties with respect to the two- and three-loop results: $\pm 1\sigma_{\als}$ (dotted)}
\label{lambda124126trans}
\end{figure}

Fig. \ref{lambda124126} shows the evolution of $\lambda$ in this framework up to the Planck scale. 
To estimate the dependence of the $\lambda$-running on the parameters $\als(M_Z)$ and $M_{{t}}$ we give the
shifted curves for $\lambda(\mu)$ when we change these parameters by $\pm \sigma$ as given in eq.~(\ref{alsmt}).\footnote{
In order not to make the plot too crowded these shifted curves are only given for the two-loop result.
The difference to the three-loop result is similar to the one between
the two- and three-loop curves for \mbox{$\als(M_Z)=0.1184$} and \mbox{$M_{{t}}=172.9$ GeV}.}
As the two- and three-loop curves are very close together we zoom in on the region where $\lambda$ crosses over to negative values
in Fig. \ref{lambda124126trans}. In this plot we give the $\als$-uncertainty for the two- and three-loop curves to compare
 between this uncertainty and the shift from two to three loops.
Note that there is a considerable difference between
\mbox{$M_H=124$ GeV} and \mbox{$M_H=126$ GeV} which
means that the evolution of $\lambda$ is very sensitive to the value
of the Higgs mass. Given a fixed value for $M_H$ the largest
uncertainty lies in the exact value of the top mass.  The second
largest uncertainty comes from $\als$. The total effect due to the
three-loop part of the $\beta$-functions is somewhat smaller than latter
as can be seen best in Fig. \ref{lambda124126trans}.
Still, it is worthy of note that   the three-loop corrections to the $\beta$-functions presented here
enhance the stability of the SM electroweak vacuum.

The smallness of the three-loop correction to $\beta_{\sss{\lambda}}$ seems to be somewhat coincidental
as  the aforementioned cancellations of individual terms in $\beta_{\sss{\lambda}}^{(3)}$  depend
strongly on the value of $M_H$.
Finding a Higgs with a mass of $124$ to \mbox{$126$ GeV} would therefore mean an excellent convergence of the
perturbation series for $\beta_{\sss{\lambda}}$.
Another intriguing consequence of a Higgs mass in that region is the
uncertainty whether $\lambda$ becomes indeed negative at high scales or not. If we take e.g. \mbox{$M_H=126$ GeV} and 
\mbox{$\als=0.1184$}
and decrease\footnote{Smaller
values for $M_{{t}}$ or larger values for $\als$ increase the stability of the vacuum.} the top mass from 
\mbox{$M_{{t}}=172.9$ GeV}
to \mbox{$M_{{t}}=171.25$ GeV} (\mbox{$M_{{t}}=171.16$ GeV} without the three-loop corrections),
then $\lambda$ stays positive up to the Planck scale $M_{Planck}$ in our framework.
The same effect can be achieved 
for \mbox{$M_H=126$ GeV} and \mbox{$M_{{t}}=172.9$ GeV} by increasing \mbox{$\als=0.1184$}
by \mbox{$6.5\,\sigma_{\als}$} \mbox{($7\,\sigma_{\als}$} without the three-loop corretions).
A combined scenario for \mbox{$M_H=126$ GeV} would be a shift of \mbox{$M_{{t}}=172.9$ GeV} 
by \mbox{$-1\,\sigma_{M_{{t}}}=-1.5$ GeV} and of \mbox{$\als=0.1184$} by \mbox{$+1\,\sigma_{\als}=+0.0007$ GeV} which would
also make $\lambda$ positive up to the Planck scale. 

Thus, we conclude that at present no definite
answer can be given to the question whether the SM vacuum is stable all
the  way up to the  Planck scale or not.
If indeed a SM Higgs boson is found with a mass of $124$ to \mbox{$126$ GeV},
this is a  good motivation for determining $\als(M_Z)$ and $M_{{t}}$ as accurately as possible
as well as calculating the SM $\beta$-functions to the highest achievable accuracy.

\section{Conclusions \label{last}}

We have computed the three-loop corrections to the evolution the
top-Yukawa coupling, the strong coupling and the quartic Higgs self-coupling
in the unbroken SM with the numerically small gauge coupling
constants $g_1$ and $g_2$ and all Yukawa couplings except for $\yt$ set to zero.

The implications of our calculation on the stability of the electroweak vacuum in the SM can be summarized as follows:
\begin{itemize}
\item
The  total effect of the three-loop terms  is relatively   small which is not
self-evident as the individual terms in $\beta_{\sss{\lambda}}^{(3)}$
are much larger than the final value due to significant 
cancellations for a Higgs mass in the vicinity of $125$ GeV.

\item  
The evolution of $\lambda$ is very sensitive to the values of the Higgs mass,
the top mass and $\als(M_Z)$.
If we
take e.g. \mbox{$M_H=126$ GeV} and decrease the top mass by about 
\mbox{$1.7$ GeV}, then $\lambda$ stays
positive up to the Planck scale (a similar observation has been made in
\cite{EliasMiro:2011aa}).  This is a very good motivation for high precision measurements
of $\als(M_Z)$ and $M_{{t}}$.
With the latter  values  known more precisely, the 
account of the the three-loop effects in the evolution of the 
quartic Higgs self-coupling would be essential in considering the
problem of the  stability of the electroweak vacuum in the SM. 
\item 
In this
context it may also be useful to calculate both the electroweak
contributions to $\beta_{\sss{\lambda}}^{(3)}$ and
$\beta_{\sss{\yt}}^{(3)}$ at three-loop level as well as the
matching of experimentally measurable on-shell parameters and
$\overline{\text{MS}}$-parameters to a higher accuracy.

\end{itemize}

We thank   Luminita Mihaila, Jens Salomon and   Matthias Steinhauser
for useful discussions and informing us on the results of  \cite{PhysRevLett.108.151602} before their publication.
We thank  Johann K\"uhn for valuable comments and support. Last but not least, 
we are  grateful to Fedor Bezrukov and Mikhail   Shaposhnikov for gently drawing our
attention to  the subject, numerous discussions  and providing  us with their version 
of a Mathematica package to perform one-loop matching in the SM.
 
In conclusion we want to mention that all our calculations have been
performed on a SGI ALTIX 24-node IB-interconnected cluster of 8-cores
Xeon computers using the thread-based \cite{Tentyukov:2007mu} version  of FORM
\cite{Vermaseren:2000nd}.  The Feynman diagrams  have been drawn with the 
Latex package Axodraw \cite{Vermaseren:1994je}.

This work has been supported by the Deutsche Forschungsgemeinschaft in the
Sonderforschungsbereich/Transregio SFB/TR-9 ``Computational Particle
Physics''.

{\em Note added.}
More elaborated analyses  of the vacuum stability which take into account
the three-loop running as obtained here and {\em two-loop} matching corrections
can  be found in  recent works 
\cite{Bezrukov:2012sa,Degrassi:2012ry,Alekhin:2012py,Masina:2012tz}.

\providecommand{\href}[2]{#2}\begingroup\raggedright\endgroup



\end{document}